\documentclass{article}
\usepackage{amsmath,amsthm,amsfonts,amssymb}
\usepackage[hang, nooneline]{subfigure}
\usepackage[autolanguage]{numprint}
\usepackage[T1]{fontenc}
\usepackage[percent]{overpic}
\usepackage{booktabs}
\usepackage[margin=0.95in]{geometry}
\usepackage{graphics}
\usepackage[normalem]{ulem}
\usepackage[toc]{multitoc}

\usepackage{multicol}
\newcommand{\np}[1]{\numprint{#1}}
\newcommand{\epssvv}{\epsilon_{\rm SVV}}
\usepackage[normalem]{ulem}

\newcommand{\new}[1]{ #1}

\usepackage{epstopdf}
\begin{document}

\title{\huge Implicit Large Eddy Simulation \\ of a wingtip vortex \\ at $Re_c = 1.2\cdot 10^6$}

\author{Jean-Eloi W. Lombard\footnote{Graduate Research Assistant, Department of Aeronautics, Imperial College, London, UK; \newline jean-eloi.lombard12@imperial.ac.uk (Corresponding Author).}, David Moxey\footnote{Research Associate, Aeronautics, Department of Aeronautics, Imperial College, London, UK}, Julien F. A. Hoessler\footnote{Senior CFD Engineer, CFD Technology, McLaren Racing, McLaren Technology Center, Woking, UK}, \\
Sridar Dhandapani\footnote{Team Leader, CFD Technology, McLaren Racing, McLaren Technology Center, Woking, UK}, Mark J. Taylor\footnote{Principle Aerodynamicist, McLaren Racing, McLaren Technology Center, Woking, UK}, Spencer J. Sherwin\footnote{Professor of Computational Fluid Mechanics, Aeronautics.}}%
\date{\vspace{-5ex}}
\maketitle

\begin{abstract}
  In this article we present recent developments in numerical methods
  for performing a Large Eddy Simulation (LES) of the formation and evolution of
  a wingtip vortex. The development of these vortices in the near wake, in
  combination with the large Reynolds numbers present in these cases, make these
  types of test cases particularly challenging to investigate numerically. We
  first give an overview of the Spectral Vanishing Viscosity--implicit LES
  (SVV-iLES) solver that is used to perform the simulations, and highlight
  techniques that have been adopted to solve various numerical issues that arise
  when studying such cases. To demonstrate the method's viability, we present
  results from numerical simulations of flow over a NACA 0012 profile wingtip at
  $Re_c = 1.2\cdot 10^6$ and compare them against experimental data, which is to
  date the highest Reynolds number achieved for a LES that has been correlated
  with experiments for this test case.  Our model correlates favorably with
  experiment, both for the characteristic jetting in the primary vortex and
  pressure distribution on the wing surface.
  The proposed method is of general interest for the modeling of transitioning
  vortex dominated flows over complex geometries.
\end{abstract}



\section*{Nomenclature }

\begin{multicols}{2}
\noindent\begin{tabular}{@{}lcl@{}}
\textit{c} &=& Wing chord \\
\textit{b} &=& wingspan \\
\textit{A} &=& aspect ratio of the wing \\
\textit{$Re_c$}  &=& Reynolds number based on wing chord \\
\textit{x,y,z} &=& Carterisan coordinates,\\
\textit{$\Delta x$} &=& local cell size \\
\textit{u,v,w} &=& Cartesian velocity components\\
\textit{$y^+$} &=& distance from surface in wall units\\
\textit{$t$} &=& time\\
\textit{$t_c$} &=& convective time associate to chord\\
\textit{$\Delta t$} &=& time-step\\
\textit{$l_0$} &=& length scale associate to largest eddies\\
\textit{$\eta$} &=& Kolmogorov length scale\\
\end{tabular}
\noindent\begin{tabular}{@{}lcl@{}}
\textit{$\tau$} &=& Kolmogorov time scale\\
\textit{$p$} &=& pressure\\
\textit{$p$} &=& far-field pressure\\
\textit{$U_\infty$} &=& inflow velocity\\
\textit{$\rho_\infty$} &=& density \\
\textit{$C_p$} &=& Pressure coefficient\\
\textit{$\nu$} &=& kinematic viscosity\\
\textit{$\alpha$} &=& angle of attack\\
\textit{$k$} &=& mode \\
\textit{$M$} &=& cut-off mode of the SVV filter\\
\textit{$P$} &=& $P=M-1$ polynomial order of the spectral element.\\
\textit{$\varepsilon_{SVV}$} &=& diffusion from the SVV filter\\
\textit{$\hat{Q}$} &=& SVV kernel

\end{tabular}
\end{multicols}

\section{Introduction}

Understanding the development and growth of wingtip vortices over lifting
surfaces is an ongoing research topic both in academia and industry. From an
academic perspective, fundamental open questions remain, such as the possible
re-laminarization of the vortex as it is shed from the wing~\cite{chow} and the
origin of meandering~\cite{Devenport:1995aa,Jacquin:2001aa,Heyes:2004}, the
low-frequency movement of the vortex core and the evolution of the vortex
structure. Vortices shed from lifting surfaces pose challenges to model in many
an industrial context such as wind turbines, helicopter blades, high-lift
configuration of aircraft and high-performance automotive
industry~\cite{Uzun:2006aa,Uzun:2010aa,Rossow1999507,Ghias:2005aa,Dacles-Mariani:1995aa}.
Developing a better understanding of the near-wake of the vortex, lying within
one chord length of the trailing edge of the lifting surface, is therefore
essential in understanding the complex flow-structure interactions of interest
in these problems. The far-field properties of these vortices are also a
challenge for the aeronautics industry, where their persistence imposes strict
limits on distances between landing aircraft~\cite{Spalart:1998}.  For these
reasons we are interested in refining modeling methods for investigating the
growth of the vortex in the near-field.

Conceptually, the simplest approach to ensure that the flow physics are
accurately simulated is to perform a direct numerical simulation (DNS), in which
all necessary scales are resolved at a given Reynolds number. For cases at even
moderately high $Re$ however, this approach is clearly unfeasible. To
demonstrate this, let us assume the Kolmogorov hypothesis holds for this flow
and as a very rough approximation that the length scale $l_0$ associated to the
largest eddies is of the same order as the chord length $l_0 \approx c$. The
 \new{number of grid points needed to resolve} the
Kolmogorov length-scale relates with the Reynolds number as
$\eta \sim Re^{-3/4}$ \new{, meaning that three-dimensional simulation of a
  uniformly turbulent flow requires a resolution of $Re^{9/4}$ grid points. For
  aeronautical test cases, where $Re$ is typically $O(10^6)$ or $O(10^7)$, we
  therefore require $O(10^{14})$ to $O(10^{16})$ grid points to resolve the
  flow.}
\new{Even accounting for variations in geometry which may permit varying
  resolution throughout the domain,} based on the current rate of advancement of
high-performance computing (HPC) facilities, resolving 
\new{fully developed three-dimensional flow at high Reynolds number} in a timely
manner will continue to be \new{well} out of reach for the foreseeable future.

\begin{figure}[ht]
 \includegraphics[trim=10pt 250pt 10pt 350pt,clip,width=1\columnwidth]{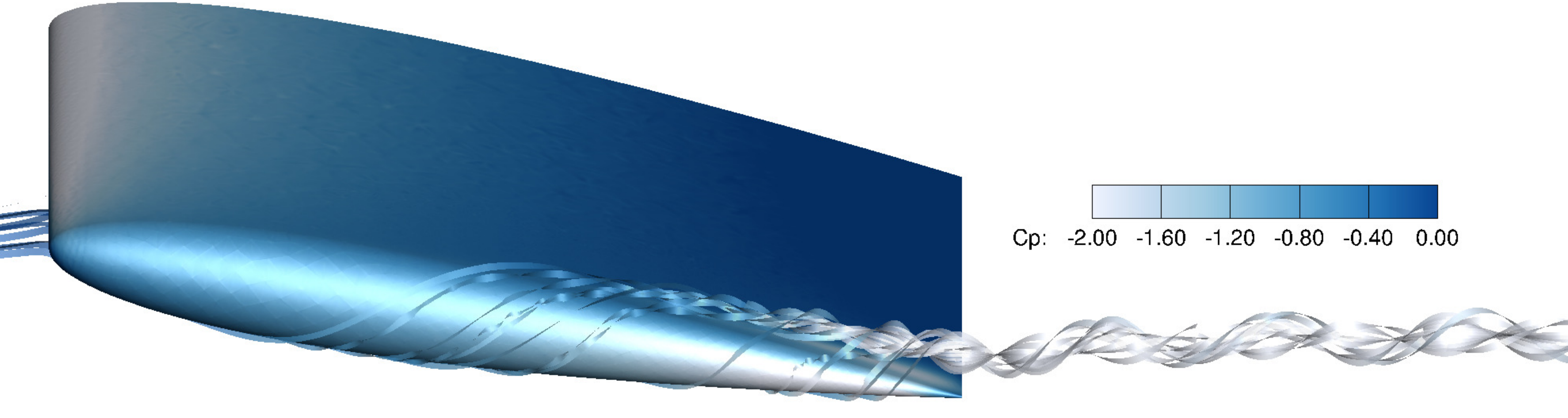}
 \caption{Wingtip vortex developing over a NACA 0012 profile with rounded wing
   cap in a wind tunnel, modeling the experimental setup of Chow \emph{et
     al.}\cite{chow} Both wing surface and streamlines are colored by static
   pressure coefficient $C_p = 2\cdot(p-p_\infty)/\rho_\infty U^2_\infty$ where
   $U_\infty^2=1$ and $\rho_\infty = 1$.
 }
 \label{f:Streamlines}
\end{figure}

Consequently, there has been ongoing development of modeling methods where small
turbulent scales are not explicitly computed. Traditional Reynolds Averaged
Navier Stokes (RANS) methods, alongside more recent advanced such as the
Reynolds Stress methods~\cite{Churchfield:2013aa}, have been developed to
simulate both the complex three-dimensional transitioning boundary layer on the
wing and the highly curved flow within the vortex. More
computationally-intensive methods, such as LES and Lattice Boltzman
VLES\cite{Satti:2012aa} have also been developed or adapted to investigate such
flows, in correlating the simulated results to
experimental data.  \new{The lattice Boltzman method has been used in
  conjunction with a modified $k-\varepsilon$ two-equation turbulence model as
  well as turbulence wall shear stress model were used to perform a VLES where
  the walls were the turbulent flow at the wall was modelled.}

\new{The key feature of these studies is in their use of reduced equations or
  turbulence models, all of which require parameters to tune their
  performance. Since the underlying physical processes that dictate the
  development and evolution of vortices is not well understood, it is therefore
  difficult \emph{a priori} to determine appropriate settings for these
  models. The aim of this work is therefore to demonstrate how an implicit LES
  method, in which the number of parameters is comparably very small and is used
  to provide additional stability, can successfully be leveraged to obtain
  accurate comparisons against experimental data.}  \new{We appreciate that
  there may be different views of the definition of implicit LES. We have
  adopted the definition of Sagaut~\cite{Sagaut:2001}, who explicitly refers to
  SVV as an implicit LES model and states that \emph{``using a numerical
    viscosity with no explicit modeling are all based implicitly on the
    hypothesis [...] the action of subgrid scales on the resolved scales is
    equivalent to a strictly dissipative action.''} The only influence of the
  sub-grid scales on the resolved scales is therefore dissipative.}

Many existing LES codes are based on finite volume, linear finite element or
Cartesian grid methods. We \new{will} instead investigate the viability of the
spectral/$hp$ element method, which lies in the class of high-order finite
element methods. These methods are widely used in academia, as they offer
attractive properties such as exponential convergence  and
low dissipation error for sufficiently smooth solutions~\cite{bolis}. This is
significant, as faithfully modeling the wingtip vortex in the far wake is of
particular interest in many research communities and industries. It is implied
that accurate modelling of the far field vortex requires precise modelling of
the vortex onset in the near field using high precision, low dissipation
schemes~\cite{Churchfield:2011aa,Dacles-Mariani:1996aa,chow,Duraisamy:2006aa,Satti:2011aa}.
However, these methods have not been widely applied for the type of industrial,
high-$Re$ cases that we consider in this work. Such methods are generally
perceived as difficult to implement, as a range of specialised preconditioners,
mesh generation procedures, \new{parallel} communication strategies
and stabilisation approaches are needed to successfully complete a simulation.

In this paper we present the SVV-iLES formulation, which utilises the spectral
vanishing viscosity approach to stabilise numerics~\cite{Tadmor:1989aa}. We show
how the issues of implementation and mesh generation can be overcome, as well as
highlight the benefits that these schemes can have for industrial problems, both
in terms of resolution power relative to existing studies and computational
efficiency. To demonstrate the viability and robustness of the scheme, we
consider the test case presented by Chow~\emph{et al.}~\cite{chow}, in which the
flow over a NACA 0012 wingtip \new{has been} investigated with
precise experimental measurements. This case has subsequently become a benchmark
for vortex dominated flows. To this end we perform an SVV-iLES, at the highest
Reynolds number considered so far for this case, and correlate our results to
the experimental data.  \new{As in previous numerical studies, in order to
  reduce the computational cost we have chosen to run the computation at a lower
  Reynolds number. In this work we set $Re=1.2 \cdot 10^6$, as compared to the
  experiment which uses $Re=4.6 \cdot 10^6$.}

In the following section we will briefly report the experimental and numerical
methods that have been developed and evaluated for investigating the nascent
wingtip vortex, as well as the key findings \new{of these studies}. We then
discuss the key features of the SVV-iLES method in Section~\ref{sec:Comp} and
the numerical methodology for our simulations. The evaluation of the SVV-iLES,
by close comparison with the extensive data from the experiment of Chow \emph{et
  al.}\cite{chow} and numerical results of Uzun \emph{et al.}\cite{Uzun:2006aa},
is presented in Section~\ref{sec:Results} before we conclude in
Section~\ref{sec:Conclusion} with a brief overview of the key findings.

\section{Literature review}
\label{sec:Lit}

We begin with a brief presentation of existing work in the investigation of
wingtip vortices; other thorough reviews of this field can be found in
Rossow~\cite{Rossow1999507}, Spalart~\cite{Spalart:1998} and Green \&
Acosta~\cite{Green:1991}. Firstly, experimental studies are considered and a
summary of the types of flow dynamics that exists for these cases is
presented. \new{We then discuss results obtained by numerical simulations before
  emphasizing the possible drawbacks of an explicit subgrid-scale model in the
  context of complex flows, such as the wingtip vortex and highlight the
  originality of the method employed in this study.}

\subsubsection{Experimental methods}

For the wingtip vortex the Reynolds number is computed from the chord length $c$
as \makebox{$Re_c = cU_\infty /\nu$}, where $U_\infty$ is the free stream
velocity and $\nu$ is the kinematic viscosity.
Giuni~\cite{giuni:2013,Giuni:2013aa} and Giuni \& Benard~\cite{Giuni:2011aa}
investigated the initial formation and development of the wingtip vortex on a
NACA 0012 rectangular wing with both square and rounded wingtips at
$Re_c = 7.4\cdot 10^5$ for angles of attack $\alpha = 0^\circ, 4^\circ$ and
$12^\circ$. They reported that for a given fixed angle of attack, a `more'
axisymmetric vortex sheds from the rounded wingtip with stronger vorticity within
its core compared the square-tip wing. They also reported that for the
$\alpha = 12^\circ$ case the axial velocity excess is higher for the rounded
wingtip. This region is surrounded by a region of axial velocity deficit
corresponding to rolling up of the vorticity sheet. The weaker axial velocity
excess for the square wingtip is believed to be a consequence of the more
numerous secondary vortices generated by the square wingtip.

Defining the center of the primary vortex as the location of the streamwise
helicity peak, they investigated the meandering of the vortex in the near
field. Two distinct modes of behaviour were observed: for the squared wingtip,
the meandering decreased with the distance from the trailing edge whereas for
the rounded wingtip case, the meandering grew quasi-linearly with distance from
the trailing edge. This has also been reported by Devenport~\emph{et
  al.}~\cite{Devenport:1995aa} and Giuni~\emph{et
  al.}~\cite{Giuni:2013aa}. However Jacquin~\emph{et al.}~\cite{Jacquin:2001aa}
concluded vortex meandering to be insensitive to free stream unsteadiness in the
wind tunnel. Fabre \&~Jacquin~\cite{Fabre:2000,Fabre:2002} and Dieterle~\emph{et
  al.}~\cite{Dieterle:1999} suggested meandering might also be influenced by the
destabilization of a vortex by secondary vortex of opposite sense. Jacquin goes
further by reporting that the various co-operative interactions between primary
and secondary vortices affected the same range of frequencies as
meandering. Zuhal~\cite{zuhal:2001} and Zuhal \& Gahrib~\cite{Zuhal:2001aa}
investigated the correlation between number and strength of secondary vortices
and the amplitude of meandering. McAlister~\cite{McAlister:1991} investigated
the wingtip vortex of NACA 0015 with both rounded and square wing cap reporting
the maximum azimuthal velocity of the vortex to be independent of Reynolds
number but dependent on the angle of attack.

Finally, Chow~\emph{et al.}~\cite{chow} performed an extensive experimental and
numerical study of the wingtip vortex over a rounded NACA 0012 profile in view
of defining a benchmark for numerical models. They reported merging of the
secondary and tertiary vortices into the primary within one chord length of the
trailing edge. From this distance onwards, the vortex had an axisymmetric
structure with a jet-like axial velocity profile (i.e. the axial velocity within
the vortex core was greater than the freestream velocity). The peak jetting
velocity was measured to be $1.77U_\infty$ at the trailing edge and by
three-quarters of a chord length downstream of the trailing edge it was measured
to be $1.7U_\infty$. Pressure taps were placed on the wing-surface to assess the
state of the boundary layer, both under the primary vortex where a strong
suction region is present and at different spanwise locations, allowing for
insight into the three dimensional boundary layer. They also reported
skin-friction lines to gain understanding into the attachment and detachment
lines of the strongest vortices in the near wake, as well as low amplitudes of
meandering because of both a relatively large angle of attack of
$\alpha=10^\circ$ and measurements at relatively short distances from the
trailing edge. Devenport \emph{et al.}~\cite{Devenport:1995aa} reported
meandering amplitude growing linearly with distance from trailing edge.

\subsubsection{Numerical methods}

Presently, RANS based methods with linear eddy viscosity models, such as
$k-\omega$ SST \new{and $k-\epsilon$}, remain commonplace for industrial flow
simulation~\cite{Churchfield:2013aa}. For vortex dominated flows, and more
generally flows with strong curvature, these models may struggle due to the
largely  \new{unsteady} dynamics of the flow. Large discrepancy
with experimental data, exceeding 100\% error on $C_p$ distribution in the
vortex core, as well as drastic under-prediction of the jetting within the core
have been reported by Churchfield \emph{et al.}~\cite{Churchfield:2009ab}. These
results have motivated the development, over the past 30 years, of more complex
closure models tailored for highly curved flows such as the wingtip vortex. We
present a brief overview of this development.

Dacles-Mariani \emph{et al.}~\cite{Dacles-Mariani:1995aa} ran a
5\textsuperscript{th} order compact (i.e. 7 point stencil instead of 11) biased
upwind scheme for the advection term and second order scheme for the viscous
term. Here they underline the necessity for low numerical dissipation. They ran
a modified version of the one-equation Baldwin-Barth turbulence model where they
modified the production term to avoid overproduction of eddy viscosity in the
vortex core. They successfully captured a secondary structure and computed the
axial velocity profiles of the core to within 3\% of experiment but
under-predicted the core pressure by more than 25\%. It should also be stressed
that Dacles-Mariani \emph{et al.}~\cite{DACLES-MARIANI:1993aa} used experimental
data to setup both inflow and outflow boundary conditions for their RANS
simulation. Linear eddy-viscosity being too dissipative, Craft \emph{et
  al.}~\cite{Craft2006684} developed a non-linear eddy viscosity model
(EVM). This more advanced model still suffered from a more severe decay of the
turbulent stresses than measured experimentally.  Duraisamy and
Iaccarino~\cite{Duraisamy:2005} modified \new{the} eddy viscosity
coefficient of the $v^2-f$ turbulence model and compared their results favorably
for axial surplus compared to the baseline Spallart-Allmaras and Menter's
$k-\omega$ SST models. The wingtip vortex exhibits a peculiarity in the
turbulence structures where the stress and strain are out of phase which renders
questionable the use isotropic eddy-viscosity based prediction methods such as
(k-$\omega$, etc). Churchfield \emph{et
  al.}~\cite{Churchfield:2008aa,Churchfield:2009aa,Churchfield:2011aa,Churchfield:2013aa}
modified the Spalart-Allmaras model to account for streamline curvature and
successfully modeled the lag between the mean strain rate and respective
Reynolds stress. This \new{method} produced the best correlation
with experiment, for a RANS based method, but proved costly (relative to other
simpler RANS methods) and so their use may be restricted to flows dominated by
vortices. They also showed that without the accurate modeling of the three
dimensional boundary layer the developing vortex remained challenging to compute
accurately even for the advanced RANS models correcting for the high degree of
curvature in the flow.

In an attempt to develop increasingly robust models, more advanced numerical
methods such as Large Eddy Simulation (LES) and Very Large Eddy Simulations
(VLES) have been proposed to investigate  \new{unsteady}
features of the flow as well as aero-acoustic properties. Fleigh \emph{et
  al.}~\cite{Fleig:2004aa} developed a compressible LES to investigate far-field
broadband noise generated by the nascent vortex on a rotating wingtip at
Reynolds $Re_c = 1\cdot 10^6$ and reported computed power and thrust
coefficients for the windmill blade within 3\% of experimental
data. Ghias~\cite{Ghias:2005aa} reported a compressible LES of NACA2415 with
square tip run at $Re_c = 10^5$ where they employed a dynamic sub-grid scale
model but did not present the correlation of their data with experiment. Uzun
\emph{et al.}~\cite{Uzun:2006aa, Uzun:2010aa} numerically investigated Chow's
experiment with a compact finite differencing LES with implicit spatial
filtering at $Re_c=5\cdot 10^5$. Their results were included in the following
comparison of our simulations with experimental data. Jiang \emph{et
  al.}~\cite{Jiang:2008} reported results for a LES simulation of Chow's
experiment at the experimental Reynolds number of $Re_c=4.6\cdot 10^6$ but did
not compare the results with experiment.  \new{The lattice Boltzman method 
  was used in conjunction with a modified $k-\varepsilon$ two equation turbulence model and
   wall shear stress model  to perform a VLES. Despite good
   correlation with the experimental results of Chow \textit{et al.}~\cite{chow}
   for the suction of the vortex on the surface of the wing the method
   over-predicted the jetting phenomena within the vortex by 23\% at
   streamwise location $x/c=-0.114$ and 12\% at $x/c=0.456$.}
So far these more advanced modeling methods have generally been used for
simulations at lower Reynolds numbers~\cite{Jiang:2008,Uzun:2010aa} or have not
not been  \new{compared against} experimental data \new{in}
the way they will be in the present study~\cite{Uzun:2006aa}.

\new{\subsubsection{Motivation and contributions of present study}}

\new{Modeling unknown physics by leveraging a sub-grid scale model outside of
  its operational window can be seen as a substantial drawback for explicit
  sub-grid scale models, which tend to rely on a wide range of parameters to
  dictate their behaviour. The two-equation eddy-viscosity $k-\omega$ \emph{SST}
  turbulence model introduced by Menter~\cite{Menter:1994aa} in 1994, for
  example, relies on 9 modeling constants. Menter underlines, in this paper, the
  strong sensitivity of the resulting computed flow to variations of 5-10\% of
  these constants and further stressing \emph{``None of the available
    theoretical tools (dimensional analysis, asymptotic expansion theory, use of
    direct numerical simulations (DNS) data, renormalization group (RNG) theory,
    rapid distortion theory, etc.) can provide constants to that degree of
    accuracy.''}
  Leveraging these many parameter sub-grid scale models, such as $k-\omega$ SST
  and more recently Reynolds Stress Relaxation models~\cite{Churchfield:2013aa},
  in complex flow cases therefore requires an \emph{a priori} knowledge of the
  flow physics. This is often infeasible, particularly when complex geometries
  are present and the length scales of both flow and geometry vary
  substantially. For example, in engineering applications such as flow over a
  Formula 1 car, different sets of parameters may be required to accurately
  model the various flow dynamics that are induced by the variations in geometry
  across the body of the the car. This includes the vortex dominated flow of the
  wing-tip regions of a front wing, regions of the front wing where the flow is
  mostly two-dimensional, and the rotating wheel that impinges on the moving
  road. It may therefore be impossible to obtain values for these parameters
  that capture the desired flow features across the entire car.

  The aim of this paper is to show that regularized high order
  spectral/\textit{hp} element methods, without an explicit sub-grid model, can
  be applied to produce results that compare favorably against experimental
  data, by considering flow over a three-dimensional geometry of practical
  interest. The SVV-iLES approach, which we will present in the following
  section, requires the choice of two regularization parameters: one to dictate
  the level of artificial viscosity, and another for a cut-off wavenumber. These
  are chosen through experimentation, such that the computation does not diverge
  but do not require assumptions regarding the physics of the flow. We discuss
  this methodology in the following section before showing results of the NACA
  0012 wingtip vortex case.}

\vspace{1em}
\section{Computational Methodology}
\label{sec:Comp}
In this section, we give a brief summary of the computational methodology used
for the NACA 0012 wingtip simulations. We begin by outlining the Nektar++
spectral/$hp$ element framework in which the solver is
implemented~\cite{Cantwell:2014}. We then outline the types of regularization
that  \new{are} necessary to perform the computations and prevent the
simulation from diverging, along with the mesh generation procedures that are
used to generate a curvilinear boundary layer mesh for the geometry. Finally, we
discuss initial and boundary conditions as well as resolution requirements in
space and time for the simulations.

\subsection{Nektar++: a high-order spectral/$hp$ element framework}

High-order finite element methods often suffer from the stigma of difficulty of
implementation, which in turn means that despite their attractive numerical
properties and the ability to resolve difficult cases such as the one presented
here, they are frequently under-used. Nektar++ is a framework designed to
address this problem by providing a modern development environment for these
methods. It is highly parallel, providing a range of efficient preconditioners
and has support for a variety of solvers including the incompressible
Navier-Stokes equations. For a summary of functionality one can see for example
Cantwell \emph{et al.}~\cite{Cantwell:2014}, or for more details on the method
itself, the reference book by Karniadakis \& Sherwin~\cite{spencer}.

In the following sections, we outline the modifications made to Nektar++ in
order to  \new{adapt} the existing DNS solver,  \new{making it suitable for an} iLES approach through
regularization. Furthermore we outline the other challenges that need to be
addressed in order to more generally make 
\new{spectral/$hp$} element \new{methods} viable for these problems.

\subsection{From DNS to SVV-iLES: filtering with Spectral Vanishing Viscosity}

Running cases at high-Reynolds number on an under-resolved mesh requires close
inspection of the source of errors. The highly non-linear nature of the
underlying equations leads to a complex interactions of these errors, which when
left uncorrected leads to a diverging solution. Broadly, we have found that the
two most important aspects are:

\begin{enumerate}
  \item consistent integration of non-linear terms;
  \item artificial dissipation to prevent divergence of the flow in the presence
  of under-resolution.
\end{enumerate}

We now explain each point in more detail. The incompressible Navier-Stokes
solver used in Nektar++ directly integrates the underlying equations through the
use of an operator splitting scheme in combination with a consistent
boundary condition \new{for the pressure Poisson
  equation}~\cite{Splitting}. In this scheme, nonlinear terms are computed
explicitly at each quadrature point, which depending on the element type uses a
form of Gauss quadrature. These nonlinear terms are then multiplied
by the elemental basis functions and integrated in order to compute the $L^2$
inner product\new{, as is required in the continuous Galerkin
  formulation}. However, since Gauss quadrature will only 
\new{produce} exact values for integrals of polynomials of 
\new{degree} $O(2P)$ at a simulation polynomial order of $P$, aliasing errors
are introduced due to the quadratic nonlinearity present in the Navier-Stokes
equations. When simulations are adequately resolved, this aliasing error usually
does not affect the robustness of the simulation. However, when under-resolution
is used for implicit LES, this aliasing effect leads to a significant buildup of
error as the simulation progresses in time, and usually causes the simulation to
abruptly diverge.

\new{We note that the nonlinear terms of the Navier-Stokes equations are
  consistently integrated if the elements are straight-sided. In this case, the
  Jacobian of the mapping which defines the coordinates of the element is affine
  with constant determinant. However, where the element is curved, this mapping
  is an isoparametric polynomial expansion, which when incorporated into
  integrands, leads to an additional source of aliasing error. In this work, we
  do not take this source of error into account. A more detailed description of
  these aliasing errors, as well as means of suppressing them, can be found in
  \cite{Mengaldo:2014}. Whilst it is true we could likely suppress aliasing
  errors using SVV, which we describe below, it would require stronger diffusion
  together with a lower cut-off mode, leading to reduced accuracy of the overall
  solution. Additionally, since the SVV operator is anisotropic, whereas
  dealiasing is isotropic, the two do not completely overlap. Therefore a
  reduced amount of regularisation can be achieved using dealiasing.}

In regards to the second point, we first note that the energy spectrum of the
flow consists in a resolved range of wave numbers, the large eddies, and an
un-resolved range, the turbulent or dissipative scales, for higher wave numbers.
Because of this cut-off the higher wavenumber dissipative scales are not
resolved for LES. The energy build-up at high-wavenumber and the coupling
through the nonlinear term down to lower wave numbers may lead to unstable
computations at worst and erroneous energy  \new{spectra} at
best.

Spectral Vanishing Viscosity (SVV), first introduced by
Tadmor~\cite{Tadmor:1989aa} for spectral Fourier methods, aims to damp the
high-wavenumber oscillations without impeding the physics of the flow at lower
wavenumbers, thereby stabilising the simulation and preserving the accuracy of
the solution. In this approach one adds an additional reaction term of the form
\[
\epssvv \frac{\partial}{\partial x}\left(\hat{Q}\star\frac{\partial u}{\partial
    x}\right)
\]
where $\epssvv$ is a constant, $\star$ denotes the convolution operator and
$\hat{Q}$ is a kernel dictating which modes receive damping. This approach was
extended to the Navier-Stokes equations by Kirby \&
Sherwin~\cite{Kirby20063128}, Karamanos and Karniadakis~\cite{Karamanos:2000}
and has been extensively used by Pasquetti \emph{et al.}~\cite{Pasquetti:2005},
Severac and Serre~\cite{Severac:2007}, Xu \emph{et al.}~\cite{Xu:2006aa} as well
as Lamballais \emph{et al.}~\cite{Lamballais20113270}.

\new{We also stress that the oscillations stabilized by the SVV method are
  sub-element oscillations. The intrinsic nature of the spectral/\textit{hp}
  element methods results in degrees of freedom within each element.  When the
  solution field is under-resolved, it is no longer guaranteed to be
  smooth. This therefore leads to the development of spurious high-frequency
  oscillations in the polynomial representation of the solution inside each
  element, arising from Gibbs phenomena occurring between two connected
  elements. The aim of SVV stabilisation is to prevent these sub-element
  oscillations, since they may lead to oscillations occurring in neighbouring
  elements and ultimately to divergence of the computed solution.}

In the context of the simulations presented here, we introduce artificial
damping as an implicit sub-grid scale model through the Spectral Vanishing
Viscosity (SVV). \new{We appreciate that there may be different
  views of the definition of implicit LES. We have adopted the definition of
  Sagaut~\cite{Sagaut:2001}, who explicitly refers to SVV as an implicit LES
  model and states that \emph{``using a numerical viscosity with no explicit
    modeling are all based implicitly on the hypothesis [...] the action of
    subgrid scales on the resolved scales is equivalent to a strictly
    dissipative action.''} The only influence of the sub-grid scales on the
  resolved scales is therefore dissipative.}

The key point in SVV filtering is that, due to the shape of the kernel
$\hat{Q}(k)$,
\[
\hat{Q}(k) = \begin{cases}
  \exp\left(-\frac{(N-k)^2}{(M-k)^2}\right), & k > M,\\
  0, & k \leq M,
\end{cases}
\]
artificial viscosity for any mode number $k$ is only applied above a cut-off
mode $M$. For the higher modes, the total
viscosity can thus be expressed as $1/Re + \epsilon_{SVV}$.

The SVV operator was incorporated into the velocity correction scheme of the
incompressible Navier-Stokes equations inside Nektar++ by following the approach
presented by Kirby \& Sherwin~\cite{Kirby20063128}, where the elemental
Laplacian operator is convolved with the kernel $\hat{Q}$. The computational
cost is therefore negligible  \new{since the only cost involved is
  during} setup. However, we note that the addition of SVV can lead to higher
iteration counts in the conjugate gradient method used to calculate the
intermediate pressure field and perform the velocity correction. This effect can
be mitigated through the use of appropriate preconditioning
strategies~\cite{lowenergy}.

In our SVV-iLES method the parameters do not adapt automatically to the flow
(i.e. without the input of an experienced user), although methods have been
proposed to overcome this constraint by implementing an adaptive SVV diffusion
\cite{Karamanos:2000,Kirby:2002aa}. These methods however still require the same
number of parameters to be calibrated on a case-by-case basis and increase the
computational runtime cost, as the matrix systems which represent the diffusion
operator need to be rebuilt whenever either parameter is changed. We have
therefore not considered such approaches here. Also note that we do not use
spatially-variable SVV diffusion coefficient or wall functions to limit
\new{the} amount of damping near solid walls.

As reference, the values of the SVV parameters used by different groups,
including the simulations performed here, are reported in Table
\ref{t:svv_comparison}.  \new{Since we do not have a detailed \emph{a priori}
  knowledge of the flow features and, consequently, of the sub-grid flow
  physics, we have chosen our SVV parameters arbitrarily so that they ensure a
  non-diverging solution. For other simpler cases in the literature, more effort
  has been made to more rigorously quantify the use of specific SVV
  parameters~\cite{Lamballais20113270}. For the sake of comparison however, we
  report the parameters used in previous studies and stress that our choice
  aligns reasonably closely with that of previous studies. We recognise greater
  exploration is necessary for more complex cases. Research along these lines is
  presently being conducted by Moura \textit{et al.}~\cite{Moura2015}.}

\subsection{Geometry and mesh generation}

\begin{table}
  \begin{center}
    \begin{tabular}{ccc}
      \toprule
      & $M$ & $\epssvv$\\
      \midrule
      Tadmor\cite{Tadmor:1989aa} & $\tfrac{1}{3}N$ & $1/N^2$\\
      Pasquetti\cite{Pasquetti:2005,Pasquetti:2006aa,Pasquetti:2008aa} & 
      $\{ \sqrt{N},  \tfrac{1}{3}N, \tfrac{1}{2}N, \tfrac{3}{5}N \}$ & $\{1/N, 1/4N, 4/N \}$\\
      Xu\cite{Xu:2006aa} & $\{N-2, \tfrac{2}{3}N\}$& $1/N$ \\
      Karamanos\cite{Karamanos:2000} & $\tfrac{15}{21}N$ & 1/N \\
      Kirby\cite{Kirby:2002aa} & $5\sqrt{N}$ & $5/8$\\
      {\bf Present study} & $\mathbf{0.5N}$ & $0.1$\\
      \bottomrule
    \end{tabular}
  \end{center}
  \caption{Different values for the SVV parameters (diffusion $\epssvv$ and
    cut-off mode $M$) where $N=P-1$ for a discretization of polynomials of order
    $P$.}
  \label{t:svv_comparison}
\end{table}

\begin{figure}
  \centering
  \includegraphics[width=0.99\columnwidth]{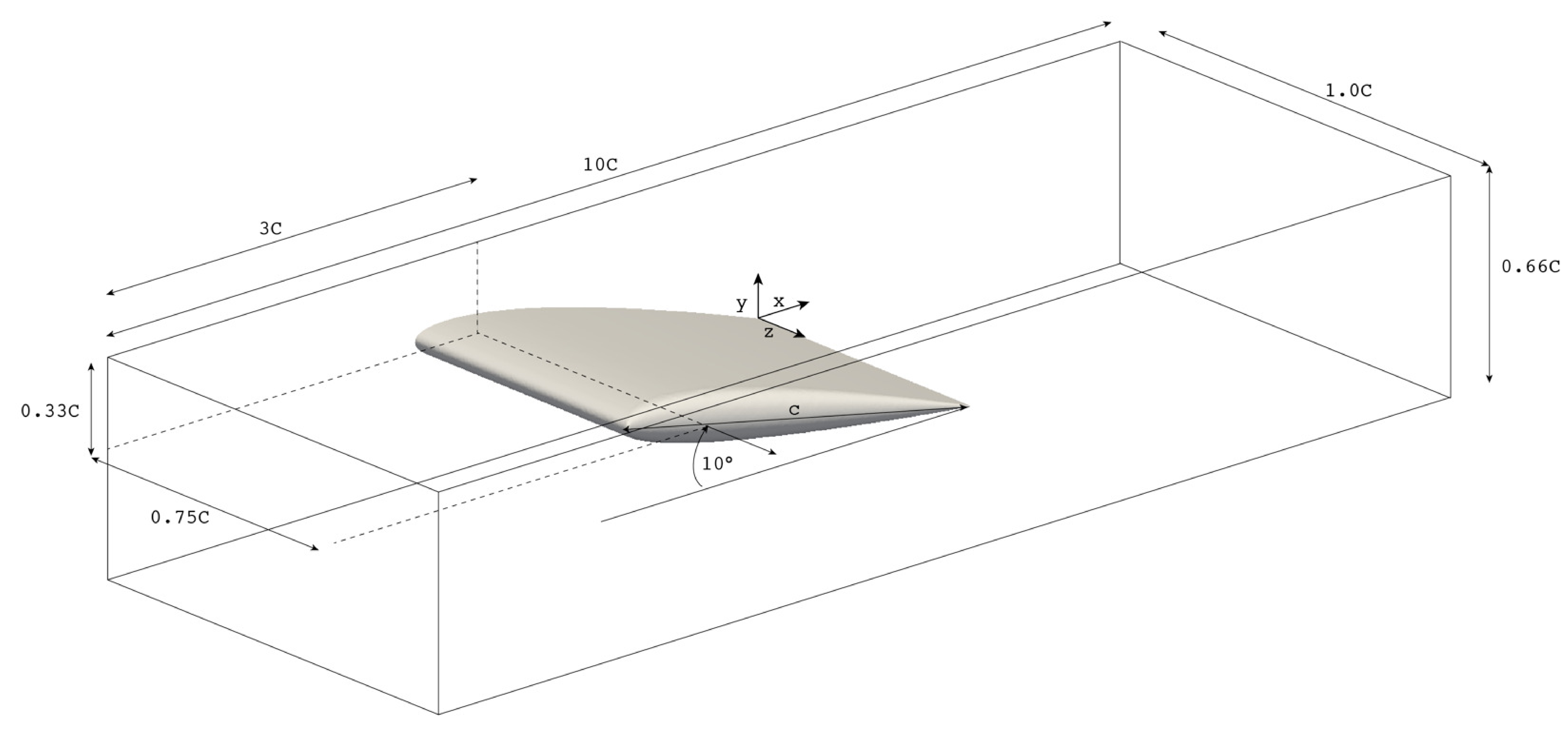}
  \caption{Computational domain representing the test section in the wind tunnel
    used by Chow \emph{et al.}~\cite{chow} for their experiment.}
  \label{fig:domain}
\end{figure}

The rectangular wing investigated numerically has a NACA 0012 profile, with a
rounded wing cap (consequently a longer semi-span where the wing is thickest)
and a blunt trailing edge. The semi-span, without the cap is, $b=0.91[m]$ and
the chord is $c=1.22[m]$ which correspond to an aspect ratio of $A=0.75$. The
boundary layer tripping mechanism~\cite{chow} used in the experiment is not
reproduced in the mesh. We represent the test section of the low speed wind
tunnel located in the Fluid Mechanics Laboratory of NASA Ames Research Center
used by Chow \emph{et al.}~\cite{chow} for the experimental work, by a
$0.66c\times c \times 10c$ cuboid domain as shown in figure \ref{fig:domain}. We
do not model the wind tunnel sections upstream and downstream of the test
section. A Cartesian coordinate system $(x,y,z)$ is used to locate point within
the computational domain with its origin the wingtip trailing edge, where $x$ is
aligned with the streamwise direction and $y,z$ are the two transverse
ordinates.

Regarding the interference between the primary vortex and the wind tunnel walls,
Chow \emph{et al.} decided to use as large a model as possible while still
avoiding severe viscous interaction between the wall and the primary vortex
through significant growth and/or separation \new{of} the boundary layers on the
wind tunnel walls. The authors of the experiment warn against large inviscid
effects (mirror effects) due to the close proximity of the wall that most likely
influence both the primary and secondary vortices. It should also be noted that
the significant blockage created by the large wing in the relatively small wind
tunnel section accelerates the flow around the wing. Hence the absolute angle of
attack, perceived by the wing, is around $2^\circ$ higher than the angle of
attack prescribed by the geometry.

\paragraph{Meshing methodology} When considering complex geometries, even
generating a linear, straight-sided mesh poses a significant challenge. We have
therefore turned to commercial mesh generators, which provide a robust approach
to generating linear straight-sided meshes,  but generally lack support for
high-order elements. We note that for high-order simulations, elements which lie
on the boundary must be curved so that they align with the underlying
geometry. Using straight-sided high-order elements can significantly alter the
physics that form near boundaries and thus downstream of the boundary.

The commercial software we use first imports the CAD geometry, in this case the
wing, and makes a fine tessellation of the surface. This tessellated surface is
then used to produce the surface mesh. As we have the meshed surface and the
original IGES (or CAD) geometry, but not the intermediary tessellation, we are
presently unable to add the necessary curvature to the linear mesh by
interrogating the CAD geometry directly. To smooth the mesh we adopt an
alternative, patch-based technique known as spherigons~\cite{volino-1998} which
rely on surface normals that are obtained from a fine triangulation generated by
the commercial software.

\begin{figure}
  \centering
  \includegraphics[width=0.6\textwidth]{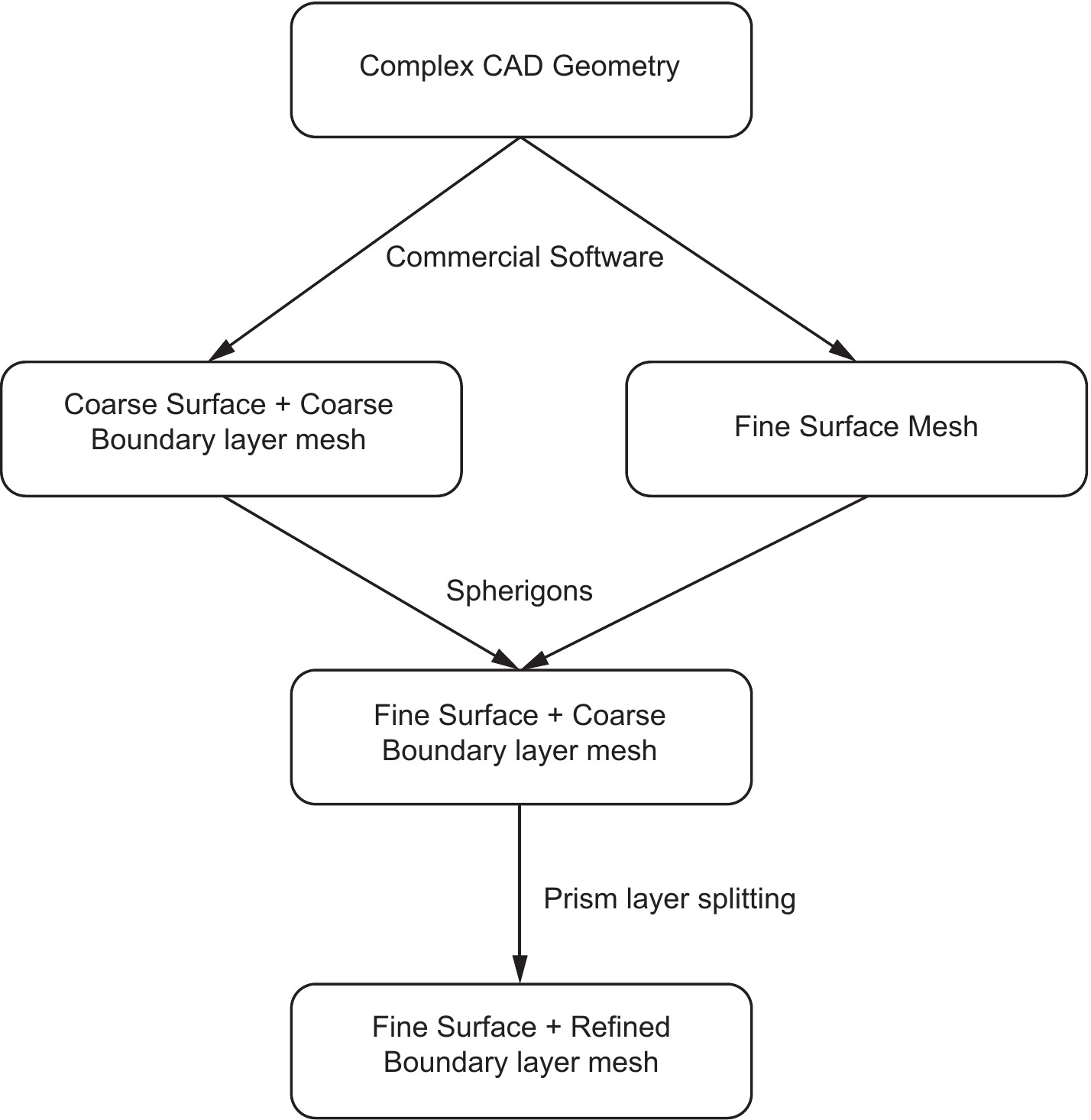}
  \caption{Robust three-step meshing procedure.}
\label{fig:3}
\end{figure}

The robust high-order three step mesh generation procedure used is summarized in
Fig. \ref{fig:3} and the interested reader should refer to Moxey
\emph{et al.}~\cite{Moxey:2014} for further details. Our meshing procedure
generates a coarse single-element boundary layer of prisms which are curved
using spherigons at the wing surface, with straight-sided tetrahedra filling the
rest of the volume. Each prism is split using an isoparametric method in the
wall-normal direction, which allows us to achieve the desired wall normal
resolution whilst preventing the self-intersection of the boundary layer
elements. The tetrahedra grow from the prism layer in a controlled manner in
regions of interest, such as the vortex path, where we impose a regular element
size to avoid introducing additional mesh induced error.

\paragraph{Comparative degrees of freedom} The implicit diffusion from the SVV
operator preserves the convergence properties of the underlying scheme. It does
not degrade the exponential rate of convergence of the accuracy of a solution
achievable for a sufficiently smooth field~\cite{Xu:2006aa}.
\new{We have focused on using our anisotropic prism refinement technique to
  capture the wall-normal near-wall resolution. In this region we have measured
  that across the wing surface, the placement of the closest grid point
  satisfies $y^+<1$. With this level of resolution, we can accurately capture
  the sub-viscous layer of the flow field and resolve the high shear of the
  boundary layer profile. However the resolution of subsequent elements is not
  enough to, for example, capture the extremely fine scale of the turbulent
  boundary layer characteristics. Additionally, for computational reasons, we
  clearly cannot resolve with a similar level of accuracy in the wall-tangential
  direction. We do appreciate that this may very well play an important role in
  capturing the flow in the region of the vortex roll-up. We discuss this
  further in the presentation of our results.}

The present mesh is composed of \np{243000} elements of which \np{24500} are
prisms around the wing surface and \np{218500} are tetrahedra growing from the
three prism layers to the wind tunnel walls. The prism layer represents roughly
20\% of the total number of degrees of freedom.  Running this computation with
5\textsuperscript{th} order polynomials (i.e. 6\textsuperscript{th} order
accuracy in space) amounts to roughly \np{16.7} million degrees of freedom;
around 40\% fewer degrees of freedom than used by Uzun \emph{et
  al.}~\cite{Uzun:2006aa} for their LES. Table~\ref{t:scheme_comparison}
compares degrees of freedom for our SVV-iLES method with other studies of this
case.

\begin{table}
  \begin{center}
    \begin{tabular}{lc}
      \toprule
      Method &  DOF ($\cdot 10^6$)\\
      \midrule
      RANS (modified Baldwin-Barth)\cite{Dacles-Mariani:1997aa}  & $2.5$ \\
      RANS (Lag RST)\cite{Churchfield:2013aa}  & $13.8$\\
      LES \cite{Uzun:2006aa} & $26.2$\\
      iLES \cite{Jiang:2008} & $26$\\
      {\bf Present study} & $\mathbf{16.7}$\\
      \bottomrule
    \end{tabular}
  \end{center}
  \caption{Comparison of mesh used by different methods, converted to degrees of
    freedom (DOF)}
  \label{t:scheme_comparison}
\end{table}

\begin{figure}
  \centering
  \includegraphics[width=0.8\columnwidth]{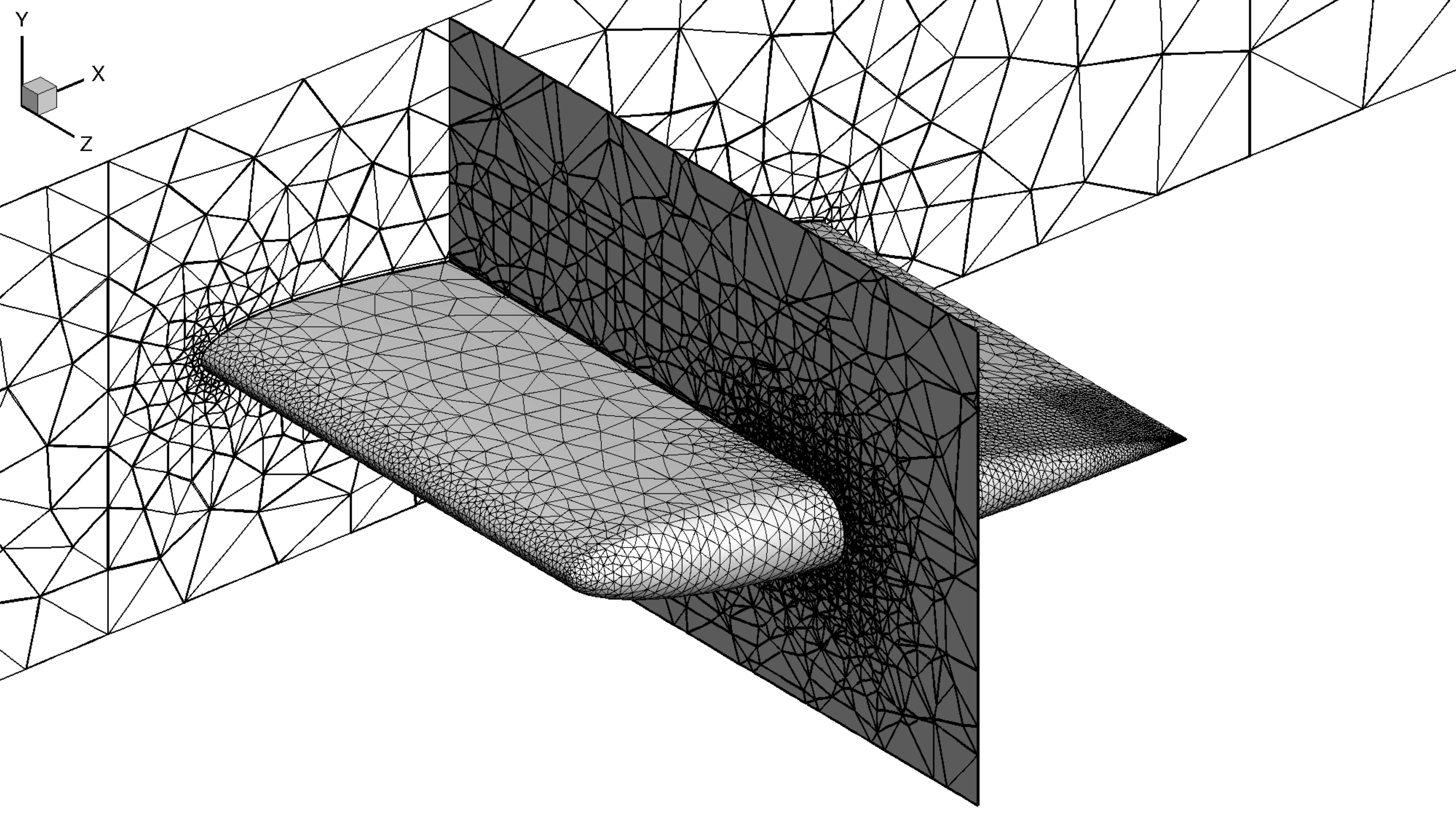}
  \caption{Overview of the coarse surface mesh and two planes, spanwise
    cut at the root of the wing and streamwise cut around mid chord. For sake of clarity quadrature/collocation points within each
    surface element are not shown here.}
  \label{fig:Mesh}
\end{figure}

\subsection{Initial and boundary conditions}

The simulation is impulsively started from $u/U_\infty =1$ throughout the
domain, except at the no-slip boundaries where $u/U_\infty = 0$, at a low
$Re_c \approx 10$. We then gradually increase the Reynolds number by a factor of
10, each time waiting for two convective time units defined as
$t_c = c / U_\infty$, until the simulation reaches the desired Reynolds number.
To improve computational efficiency the polynomial order $P$ of the
discretization is increased with Reynolds number, with $P=2$ for 
$Re_c = 10$ and $P=6$ for $Re_c=1.2\cdot 10^6$. At the outflow, we impose the
boundary condition developed by Dong \emph{et al.}~\cite{Dong201483} that
balances the kinetic energy influx through the outflow boundary condition to
prevent instability. The computational setup differs from the experimental setup
in three ways. We discuss these and their possible influence on the results in
the following paragraph.

Firstly, the boundary layers developing on the wind tunnel walls are neglected,
by using a free slip condition. The primary vortex is located sufficiently far
away so that the viscous interaction between the primary vortex and the wind
tunnel walls is much weaker than the interaction between the primary vortex and
the wing surface via the secondary structures.
\new{As a first approximation, the wall acts, inviscidly, as a symmetry
  condition. Since the computed location of the vortex is similar to the
  experimental data, the inviscid interaction between the vortex and the
  wind-tunnel walls is assumed to be of similar intensity.}
Secondly, as with other LES studies, we do not model turbulence at inflow. As we
shall see in the results section, despite the lack of a turbulent inflow we
still see good comparisons against experimental data. Finally the boundary layer
on the wing is not tripped. The tripping of the boundary layer near the leading
edge has been reported to increase the diameter of the vortex (measured by the
peak to peak distance of the vertical velocity) by 30\%
\cite{McAlister:1991}. McAlister also report that adding a boundary layer trip
changes the streamwise component of the velocity from a small excess to a large
deficit at the position $x/c=4$; however we do not observe the vortex that far
downstream. The tripping of the boundary layer might affect the interaction
between the primary and secondary vortices and has been reported to decrease the
inboard movement of the primary vortex along the span~\cite{McAlister:1991}.

\subsection{Temporal evolution}

The Navier-Stokes equations are integrated \new{in time} using a second-order
accurate stiffly-stable implicit-explicit scheme~\cite{Karniadakis1991}. The
advection term is explicitly integrated, whereas the viscous term is implicitly
integrated. This therefore relaxes the sometimes stringent, diffusion stability
condition $\Delta t \propto \Delta x^2 / \nu$. The CFL restriction
$\Delta t \propto \Delta x / u$, where $u$ is the advection velocity within each
cell, remains however. Because the mesh we consider is coarse, we assume
numerical error is dominated by error from the spatial discretization. The time
step used for the computation normalized by the convective length scale of the
chord $t_{c}$ is $\Delta t/ t_{c}= 1.6\times 10^{-6}$ which translates into a
maximum normalized sampling frequency of $3\cdot 10^5$.  For spectral/$hp$
element methods using a second order implicit-explicit time integration scheme,
the analog to the CFL definition imposes a restriction on the maximum timestep
of the form\cite{spencer}
$\Delta t < \Delta x/(\max_{\Omega^e\in\Omega}\{|V^e| P^2\})$ where we assume
$\max|V^e| \sim U_{\infty}$. For the present mesh, where the smallest mesh
element is $10^{-4}c$ and using 5\textsuperscript{th} order polynomials, the
timestep restriction is of the order of $10^{-5}t_c$. In practice, since the
velocity in some regions can be significantly larger than $U_\infty$, we use a
timestep one order of magnitude smaller.

\section{Results and Discussion}
\label{sec:Results}

In this section, the performance of the SVV-iLES \new{method} is evaluated by
simulating the development of the wingtip vortex in the near field and comparing
the results against the experimental study by Chow \emph{et al.}~\cite{chow}
\new{as well as the previous LES by Uzun~\emph{et al.}~\cite{Uzun:2006aa}}.

We define the \emph{near field} to be the region above and below the wing and up
to one chord length downstream of the trailing edge. In this context the
\emph{mid-field} is the region from $c$ to $10c$ and the \emph{far-field} is at
a distance of more than $10c$ from the trailing edge. To put the challenge of
accurately computing the vortex into perspective, we outline the typical
tracking distances of interest for different applications. The automotive
industry is interested in tracking the vortex for roughly $20c$. For wind
turbines and helicopters blades, with an aspect ratio of roughly 10, the study
of the interaction between the rotor blades and the preceding blades that leads
to noise and structural vibration, requires tracking the wingtip vortex over
more than $60c$ for one revolution.
In this study an effort is made to characterize the performance of the modeling
method both in the three-dimensional turbulent boundary layer and in the rollup
wake. It is supposed that accurate modeling of the vortex in the near field
downstream of the trailing edge pre-supposes an accurate modeling of the
three-dimensional boundary layer roll-up on the wing surface.

\begin{figure}
  \centering
  \includegraphics[width=0.99\columnwidth]{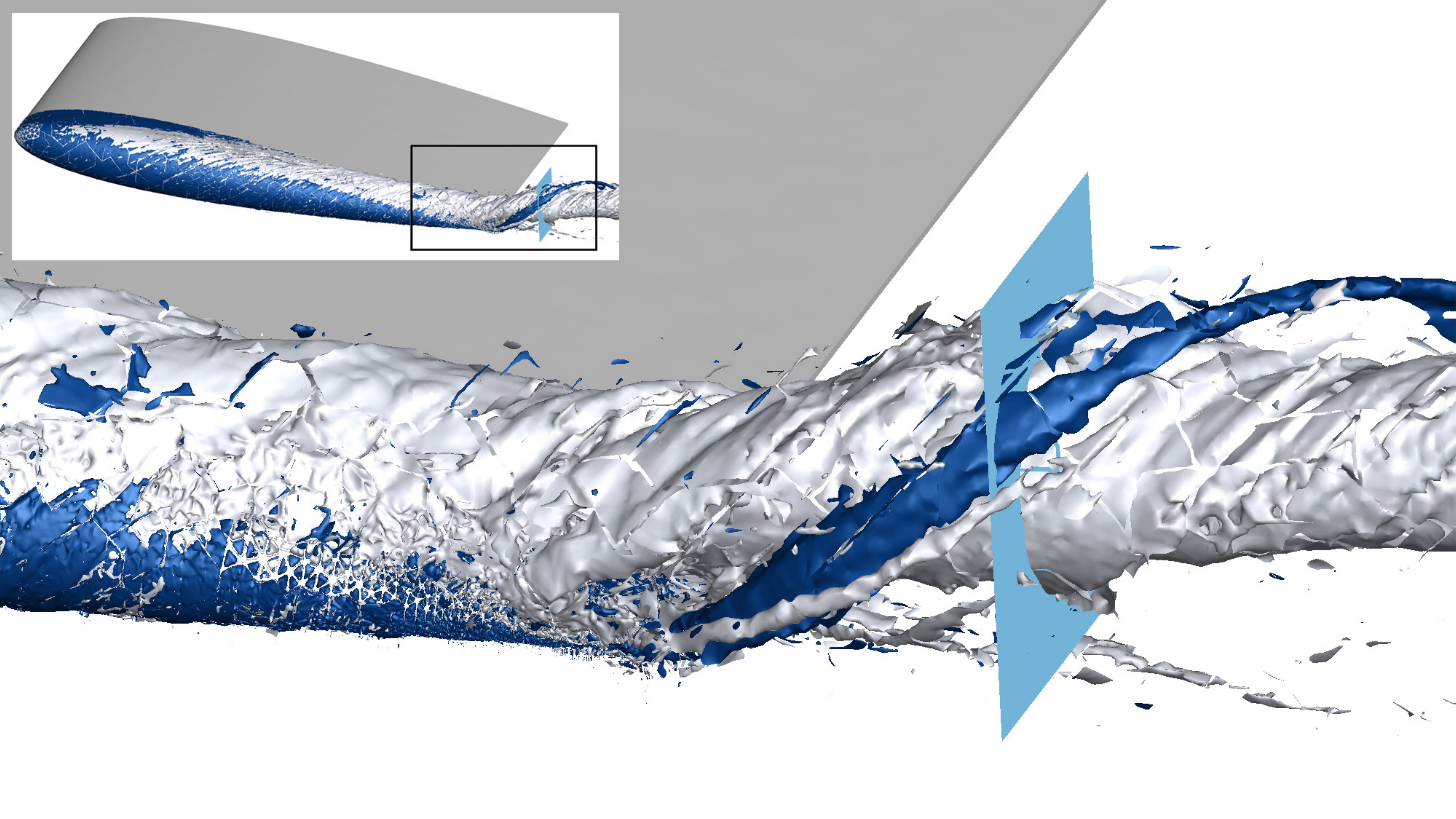}
  \caption{Contours of iso-helicity showing the interaction
    between the primary (in light grey) and a secondary vortex of opposite rotational
    sense (in blue, or dark grey). The light grey/blue plane denotes the location of the streamwise cut $x/c=0.125$ at
    which results are compared with experiment in Figs.~\ref{fig:10}-\ref{fig:11}.}
  \label{fig:7}
\end{figure}

The vortex can loosely be defined as the region in which the fluid has high
 \new{helicity}, low relative pressure and an axial velocity surplus. Different methods
have been developed for defining the center of a vortex, but these can prove
challenging to apply in the near wake where the vortex is forming as a
consequence of the shear layer roll-up. These methods include helicity peak
correction, vorticity peak, $Q$-criterion, zero in-plane velocity and axial
velocity peak. It has been shown that the helicity peak correction method is
best suited for estimating the vortex core radius, axial velocity peak and swirl
velocity peak.  Giuni \& Benard~\cite{Giuni:2011aa} assert that the centering
method chosen should depend on the key aspect of interest but the robustness of
these methods is not sufficient for identifying the vortex as it develops from
the shear layer roll-up and interacts with secondary structures. For this reason
the method of the manual location of the vortex core by identifying local
pressure minima has been favored. This method has also been used by
Chow~\emph{et al.} so this source of error should be taken into account when
appreciating the discrepancy between reported core locations
(Fig. \ref{fig:8}) in this region.

The flow over the wingtip develops into a highly skewed three-dimensional
boundary layer that rolls up and detaches into a rapidly rotating vortex, at a
distance of around $0.5c$. This forms an increasingly low pressure region in the
vortex core that gradually accelerates the fluid entering the core into a jet
characterized by a notable normalized axial velocity surplus. This strong vortex
is thought to be laminar and persistent, extending \new{many chord lengths} downstream
of the trailing edge. The challenge from a modeling perspective comes from the
three-dimensional boundary layer, the detachment and the strong curvature
induced both by the geometry and by the many interacting vortical structures.
Two key regions of the flow are used to assess the performance of the SVV-iLES
method against the experimental data of Chow \emph{et al.}~\cite{chow}: the wing
surface and the vortex core. These two regions are of particular interest
because they radically differ in nature. Most classical turbulence models can
capture turbulent boundary layers well, but few of these can accurately model
the vortex core where curvature of the streamlines is high~\cite{Bradshaw:1973}.

In the second region of interest, the vortex core, the key feature to reproduce
is the low pressure region driving the jetting phenomena or the normalized axial
velocity surplus. Inside this region, the high value of $u_{\max}/U_{\infty}$ at the
trailing edge, and the low decay within the first chord length downstream of the
trailing edge at $x/c=0.867$ is a challenging feature to capture. \new{We will
  therefore compare our results, as well as those from previous computations,
   against the reported experimental values of
  $u_{\max} = 1.77U_\infty$ and $1.59U_\infty$ respectively from Chow
  \emph{et. al.}~\cite{chow}.}  Uzun~\emph{et al.}~\cite{Uzun:2006aa}
underpredict the peak normalized and time averaged axial velocity by more than
20\%, albeit at a lower chord Reynolds number $Re_c = 5\cdot 10^5$. The
$k-\omega$ based RANS computation by Churchfield~\emph{et
  al.}~\cite{Churchfield:2011aa} report a 150\% error, \new{with respect to the
experimental data by Chow \emph{et. al.}~\cite{chow}}, in the $C_p$ distribution
within the vortex at $x/c=0.867$. Even the $k-\omega$ SST-RC
RANS~\cite{Churchfield:2011aa}, that predicts the $C_{p_0}$ at $x/c=0.867$ with
less than 10\% error, has nearly 20\% error for the estimation of the axial
velocity surplus and 30\% error for the static pressure in the core at
$x/c=0.867$.

The results from SVV-iLES and evaluation with respect to experiment for the two
regions of interest is presented in the next section.  All results presented here have been
time-averaged over three chord convective length or $1t_{c}$.  \new{We assume the
  flow is fully developed when the $C_p$ distribution on the wing-surface at the
  span-wise location $z/c=0.899$ converges to a smooth curve.}

\subsection{Wing Surface}
\subsubsection{$C_p$ distribution}
\begin{figure}
  \centering
    \begin{overpic}[trim=20pt 20pt 10pt 60pt,clip,width=0.45\textwidth]{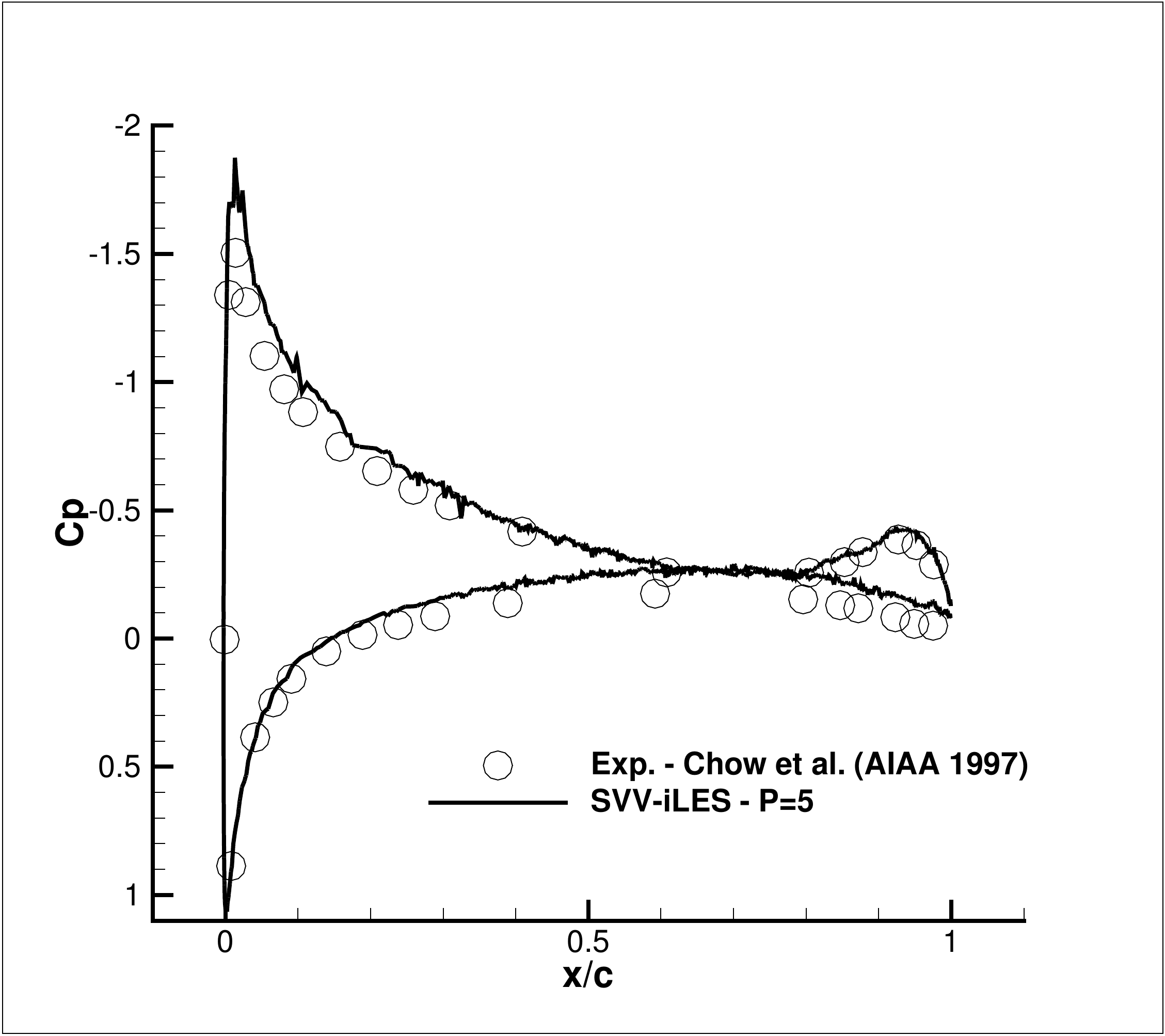}
      \put(25,75){(a)}
    \end{overpic}
    \begin{overpic}[trim=20pt 20pt 10pt 60pt,clip,width=0.45\textwidth]{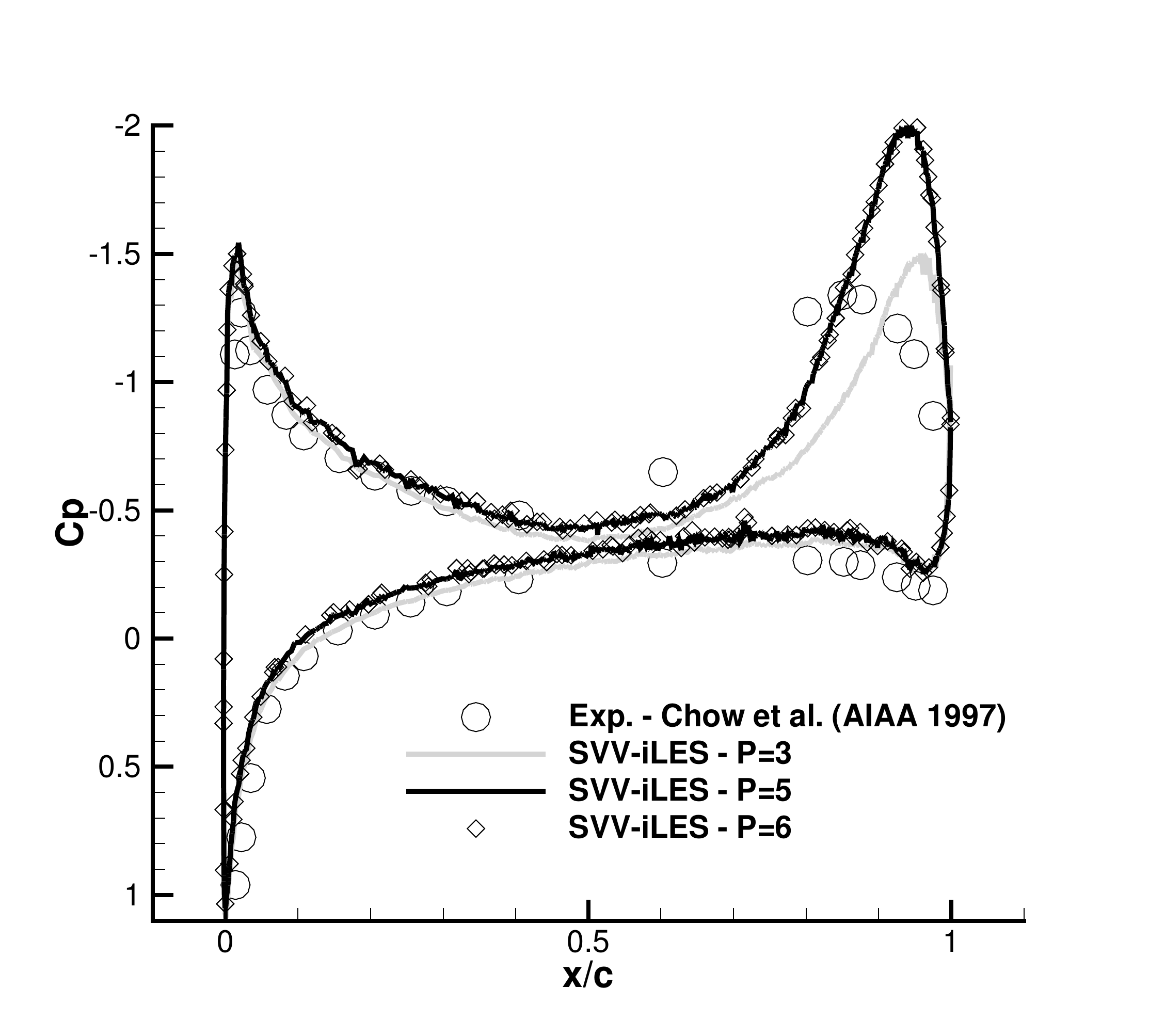}
      \put(25,75){(b)}
    \end{overpic}
  \caption{Comparison with experiment~\cite{chow} of time-averaged $C_p$
    distribution over $3t_c$ time units as a function of streamwise position,
    with the leading edge in $x=0$, at spanwise location $z/b=0.833$ (a) and
    $z/b=0.899$ in (b) where we also show the results from the uniform
    grid-refinement study (convergence in \textit{P}). The number of mesh
    degrees of freedom for 4\textsuperscript{th}, 6\textsuperscript{th} and
    7\textsuperscript{th} order accurate in space are 5.7M, 16.7M and 25.3M
    respectively. The 50\% increase in number of degrees of freedom when using
    7\textsuperscript{th} instead of 6\textsuperscript{th} does not
    significantly affect the pressure distribution on the wing at the spanwise
    location $z/c=0.899$.}
\label{fig:5}
\end{figure}

Chow \emph{et al.}~\cite{chow} reported the pressure distribution at two
spanwise locations. The first cut, at $z/b=0.833$, is situated inboard of the
vortex core where its influence is mild, whereas the second is located at the
vortex core in $z/b=0.899$. At this position, the presence of the vortex leads
to a distinctive pronounced suction region. The extent of this region upstream
depends on the shape of the roll-up layer. The presence of the vortex reduces
the pressure in this region which translates into an increase in
lift.

Despite a relatively good agreement with experiment for the $C_p$ distribution
at the spanwise location $z/c=0.833$ (Fig. \ref{fig:5}a), at the spanwise
location of the developing primary vortex (Fig. \ref{fig:5}b), the SVV-iLES
computed results under-predict the vortex suction from $x/c=0.5$ to $x/c=0.9$
and significantly over-predict the suction for the last $0.1c$. \new{Coarse
  tangential resolution, in the first 0.5c, may have led to a strong SVV
  dissipation that which in turn significantly damped the early growth of the
  vortex over the wing surface. The sudden change in trend at $x/c=0.9$ might be
  due interaction between the primary and a secondary the secondary vortex.}
With this exception however, the main features of the flow are well captured.

\new{It should be noted that the instantaneous field is noisy because of the
  unsteady nature of the boundary layer at this high Reynolds number and is
  obviously reduced as we average over longer time intervals, which explains the
  residual noise in the $C_p$ distribution (Fig. \ref{fig:5}a-b). The use of the
  spherigon mesh smoothing technique in the representation of the geometry may
  also lead to some higher frequency oscillations and thus less smoothness in
  the distribution.  Adding further diffusion from the SVV may help in removing
  some oscillations. However, additional diffusion might lead to an artificial
  reduction of the Reynolds number.}

\new{\subsubsection{Resolution study} Although this test case is computationally
  expensive to simulate, we have performed a limited $p$-refinement study of the
  flow physics, using the $C_p$ distribution as a benchmark for observing
  convergence and providing a form of self-validation. In these tests, the
  polynomial order was varied, comparing the $P=5$ results that we present here
  to results at $P=3$ and $P=6$. The resulting $C_p$ distributions, presented in
  figure~\ref{fig:5}b, show very little difference between the $P=5$ and $P=6$
  cases, despite a 50\% increase in the total number of degrees of
  freedom. However, there is a significant difference between that of $P=3$ and
  $P=5$. Whilst there is still variation between the experimental results and
  both $P=5$ and $P=6$, we can at least conclude that in terms of polynomial
  order the simulation is well-resolved.}

\new{We note that this type of refinement is good in that it is hierarchical and
  so all the degrees of freedom at lower polynomial orders are contained within
  the higher order simulations. It does not, however, guarantee that there
  cannot be regions which are still not captured and so the basic features of
  the flow will remain reasonably similar. Further work is therefore required to
  generate a larger sequence of meshes to study the effect of refinement in
  terms of element size. However, given the significant undertaking of this type
  of study and the convergence we have obtained in $p$, we do not consider this
  here and use the $P=5$ results in the coming section.}

\subsection{Propagation of the vortex core in the near-wake}
\begin{figure}
  \centering
    \begin{overpic}[width=0.45\textwidth]{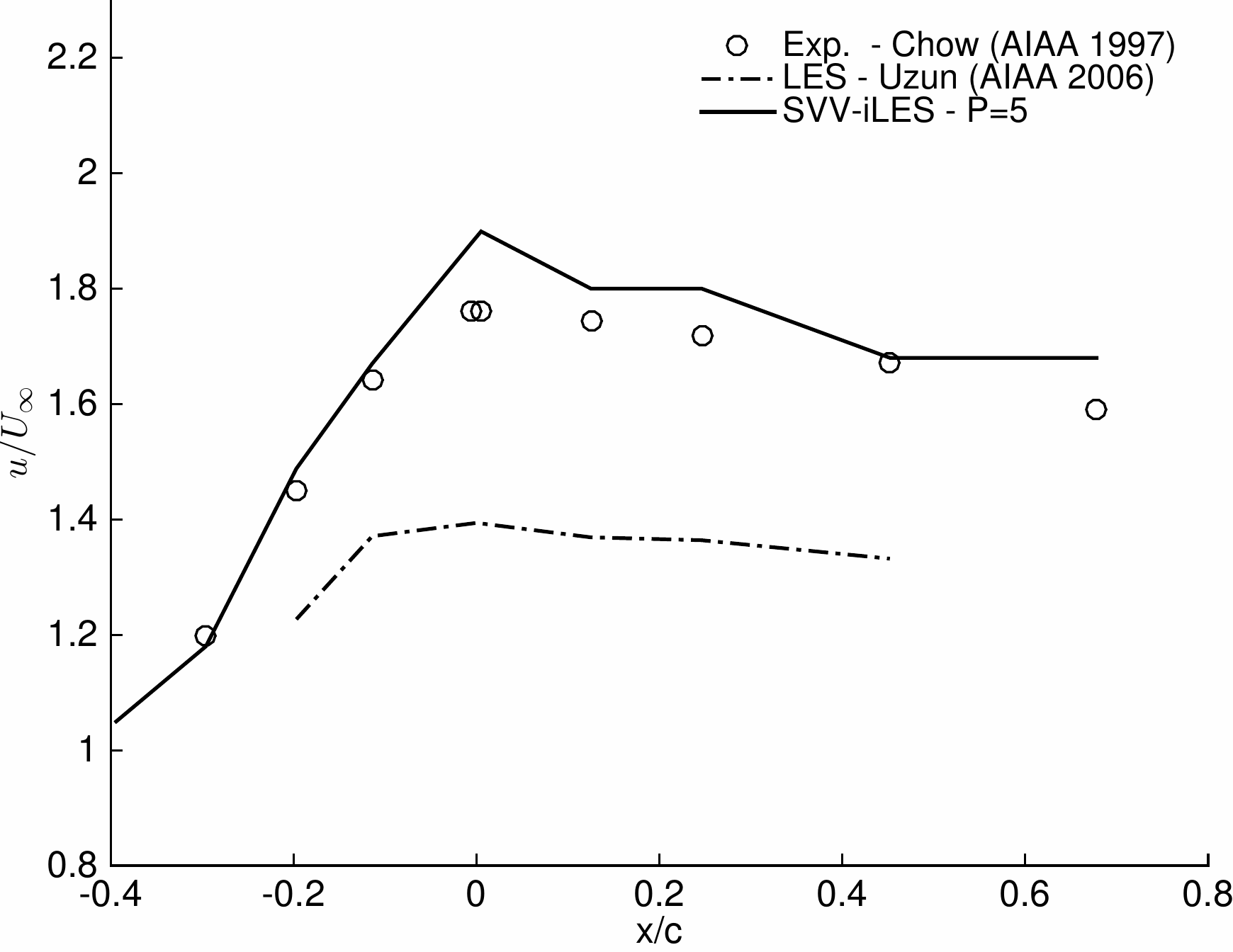}
          \put(20,70){(a)}
          \end{overpic}
    \begin{overpic}[width=0.45\textwidth]{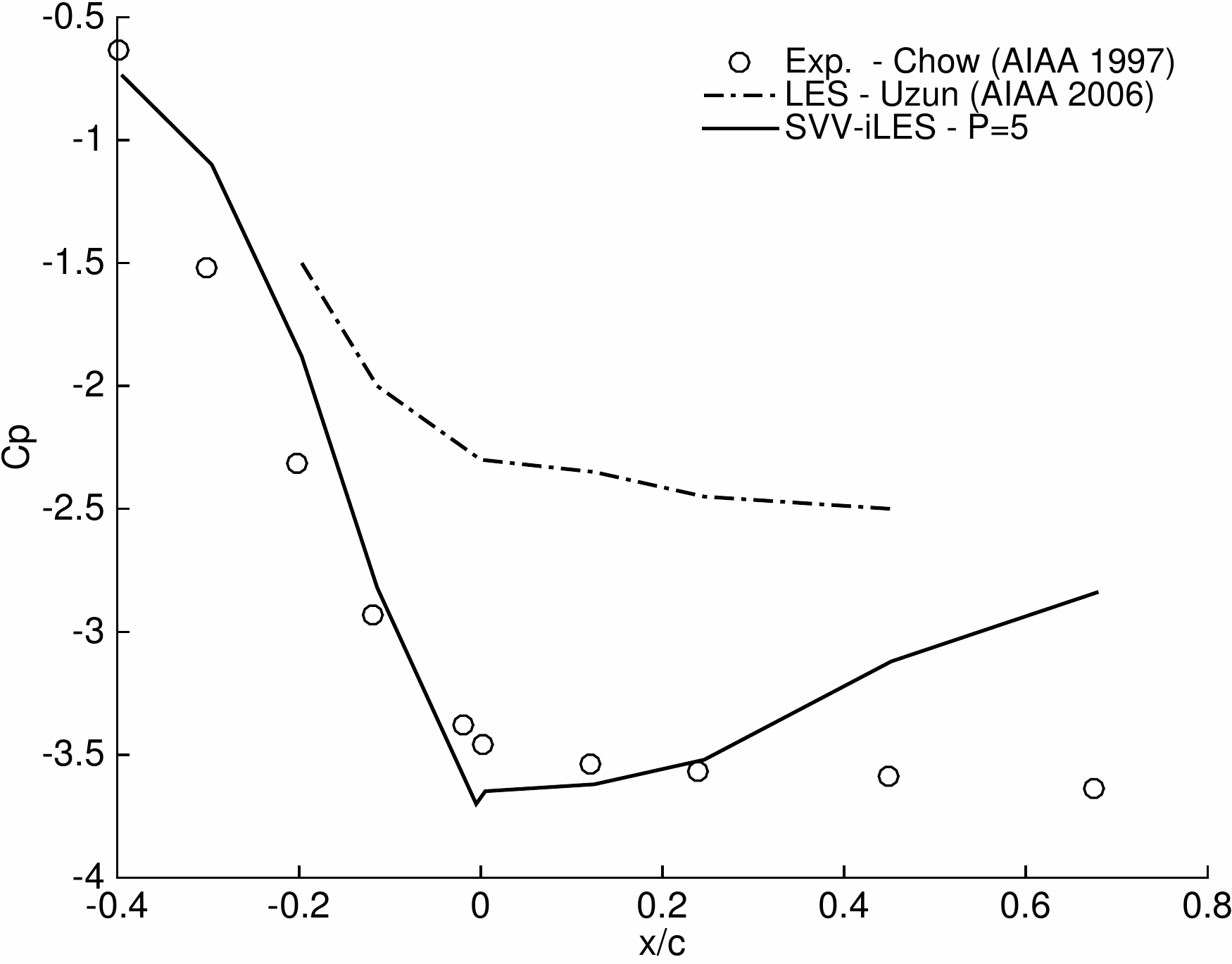}
          \put(20,70){(b)}
	    \end{overpic}
  \caption{Comparison with experiment~\cite{chow} and previous LES by Uzun
    \emph{et al.}~\cite{Uzun:2006aa} of the progression of (a) axial velocity and
    (b) $C_p$ distribution in the vortex core. The origin, \new{$x/c=0$}, is taken to be the
    position of the wingtip trailing edge.}
  \label{fig:streamwiseProg}
\end{figure}

In the vortex core we track the normalized time-averaged axial velocity, the
static pressure and at the location of the vortex both in the spanwise direction
($z$) and normal to the suction side direction (positive $y$), as shown in
figures~\ref{fig:streamwiseProg} and \ref{fig:8}. Therefore both
time-averaged pressure coefficient and axial velocity have not been corrected
for the error stemming from possible meandering. In this section we also presents an 
overview of the development of the vortex in the near
wake. A comparison shown for both streamlines and time-averaged normalized axial
velocity at two crossflow locations: over the wing towards the trailing edge at
$x/c=-0.114$ and in the near wake at the $x/c = 0.125$. These results are shown
in figures~\ref{fig:9} and \ref{fig:10}.

\subsubsection{Normalized axial velocity in the vortex core.}
The progression of the axial velocity of the vortex can also be affected by the
presence/absence of a boundary layer trip as reported by
McAlister~\cite{McAlister:1991}: for $Re_c=1.5\cdot 10^6$ a NACA 0015 profile at
angle of attack $\alpha=12^\circ$ the streamwise component of the velocity in
the vortex core has a small jetting behavior (< 5\% velocity excess) without the
trip and a significant 20\% deficit ($u/U_\infty \approx 0.8$) when a boundary
layer trip is added to the leading edge. In Fig.~\ref{fig:streamwiseProg:U} we
report the streamwise progression of the normalized axial velocity. Our modeling
adequately resolves the strong jetting behavior measured experimentally. We do
however over-predict the peak axial velocity in the vortex core by 6\% \new{with
  respect to the experimental value}.

\subsubsection{Static pressure within the vortex core.}
In Fig.~\ref{fig:streamwiseProg:Cp} we report the streamwise progression of the
vortex core static pressure. Despite a relative error of less than 10\% over the
wing surface the error grows linearly downstream of the trailing edge reaching
30\%. Accurately capturing the low pressure region within the vortex core and
sustaining this low pressure even just one chord distance downstream of the
trailing edge is particularly challenging. The linear increase in pressure is
most likely the result of either too coarse a mesh within the vortex core and/or
too strong a contribution from the SVV. \new{When the grid is too coarse to
  adequately capture a gradient the SVV filter damps out part of its kinetic
  energy. Neglecting compressibility effects, this resulting decrease in jetting
  velocity increases the pressure (or decreases the suction of the vortex).}

\subsubsection{Vertical position of the vortex core.}
\begin{figure}[ht]
  \centering
    \begin{overpic}[width=0.45\textwidth]{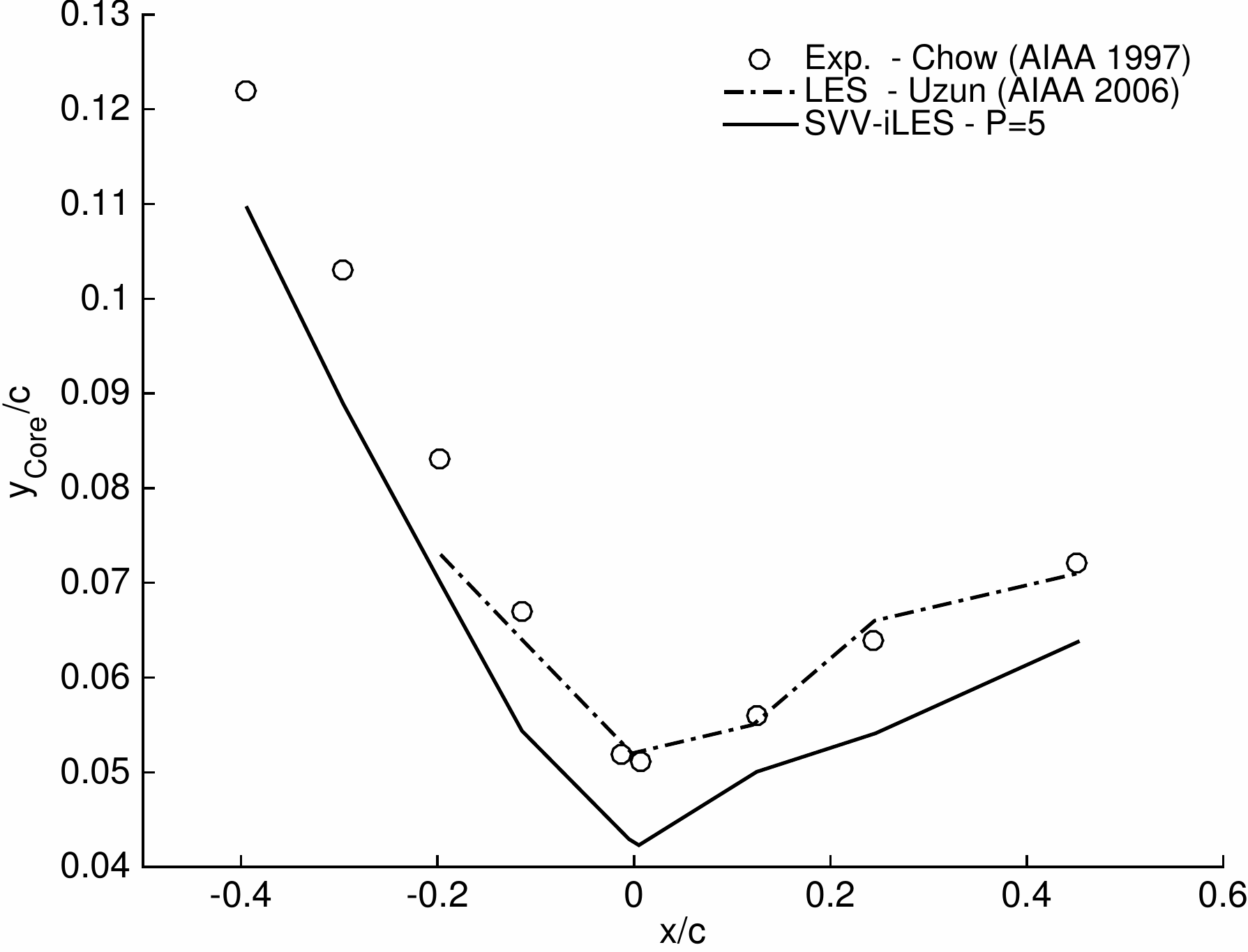}
              \put(25,70){(a)}
          \end{overpic}
    \begin{overpic}[width=0.45\textwidth]{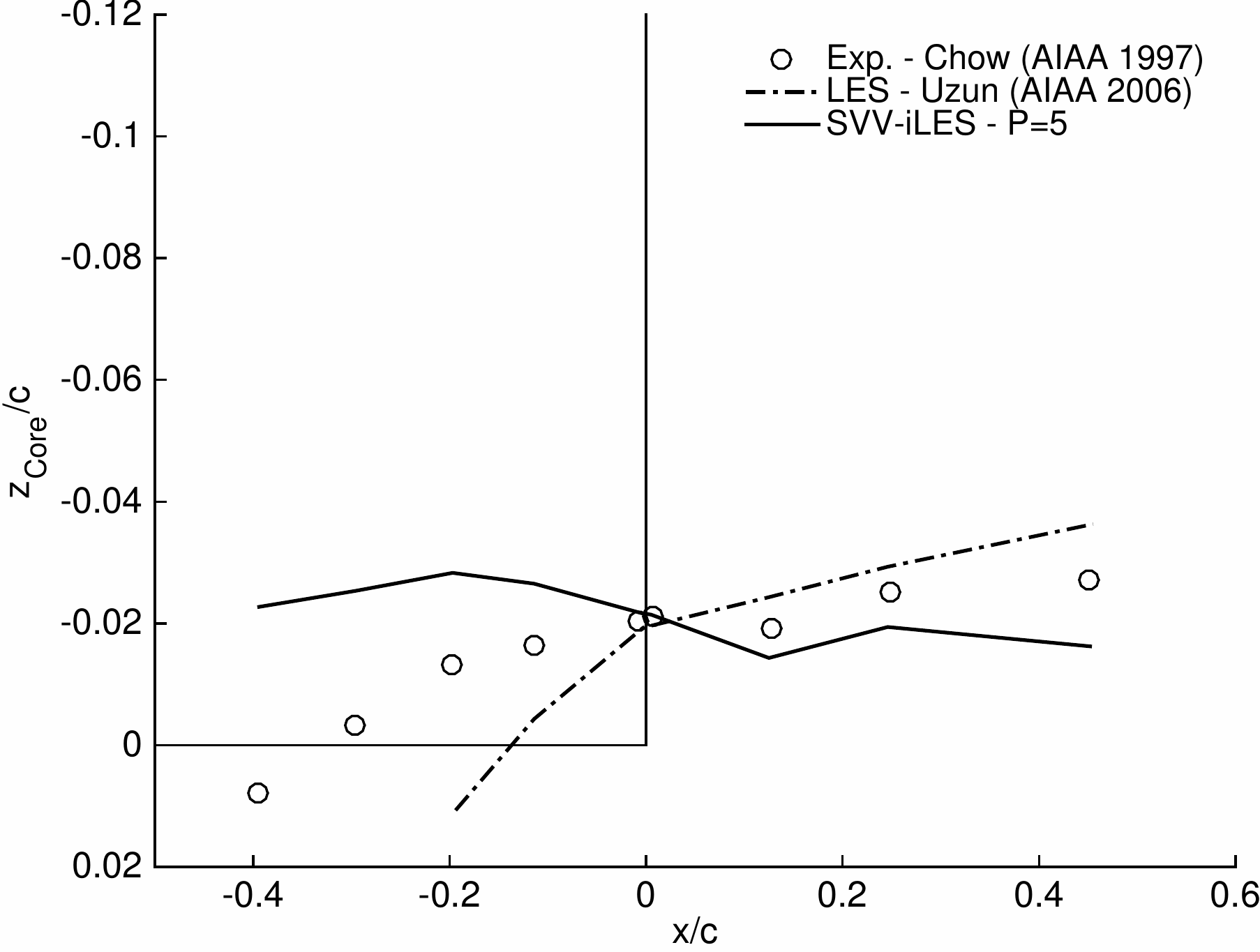}
              \put(25,70){(b)}
          \end{overpic}
  \caption{Comparison with experiment~\cite{chow} and previous LES by Uzun
    \emph{et al.}~\cite{Uzun:2006aa} of the vertical position of (a) the vortex
    above the wing and (b) spanwise location. The origin, $z_{core}/c=0$ and $y_{core}/c=0$ is taken to be the position of the
    wingtip trailing edge. The position of the wing is identified with the rectangle in b).}
  \label{fig:8}
\end{figure}

\begin{figure}[ht]
  \centering
\begin{overpic}[trim=20pt 20pt 20pt 20pt,clip,width=0.32\columnwidth]{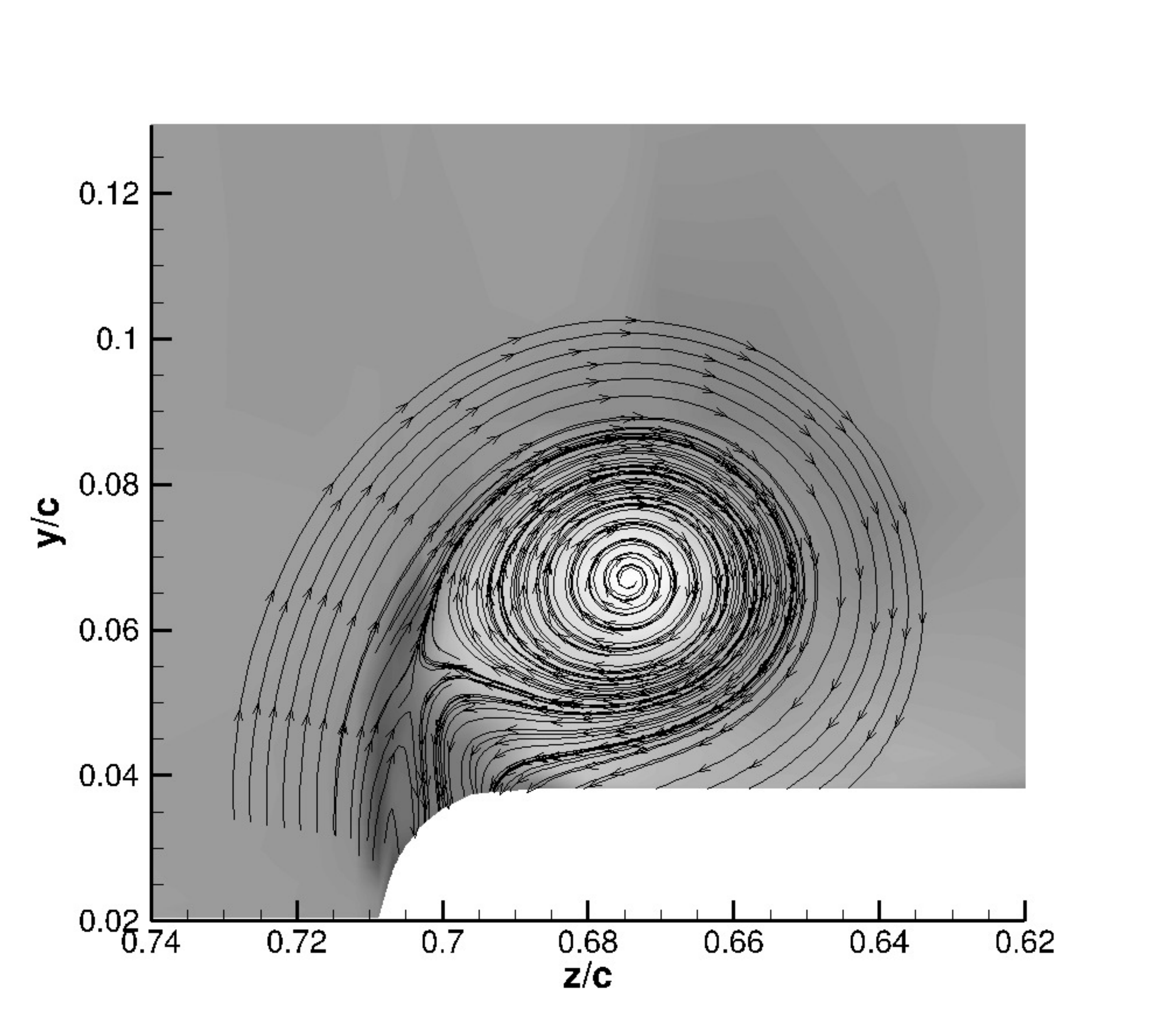}
              \put(20,70){(a)}
          \end{overpic}
\begin{overpic}[trim=20pt 20pt 20pt 20pt,clip,width=0.32\columnwidth]{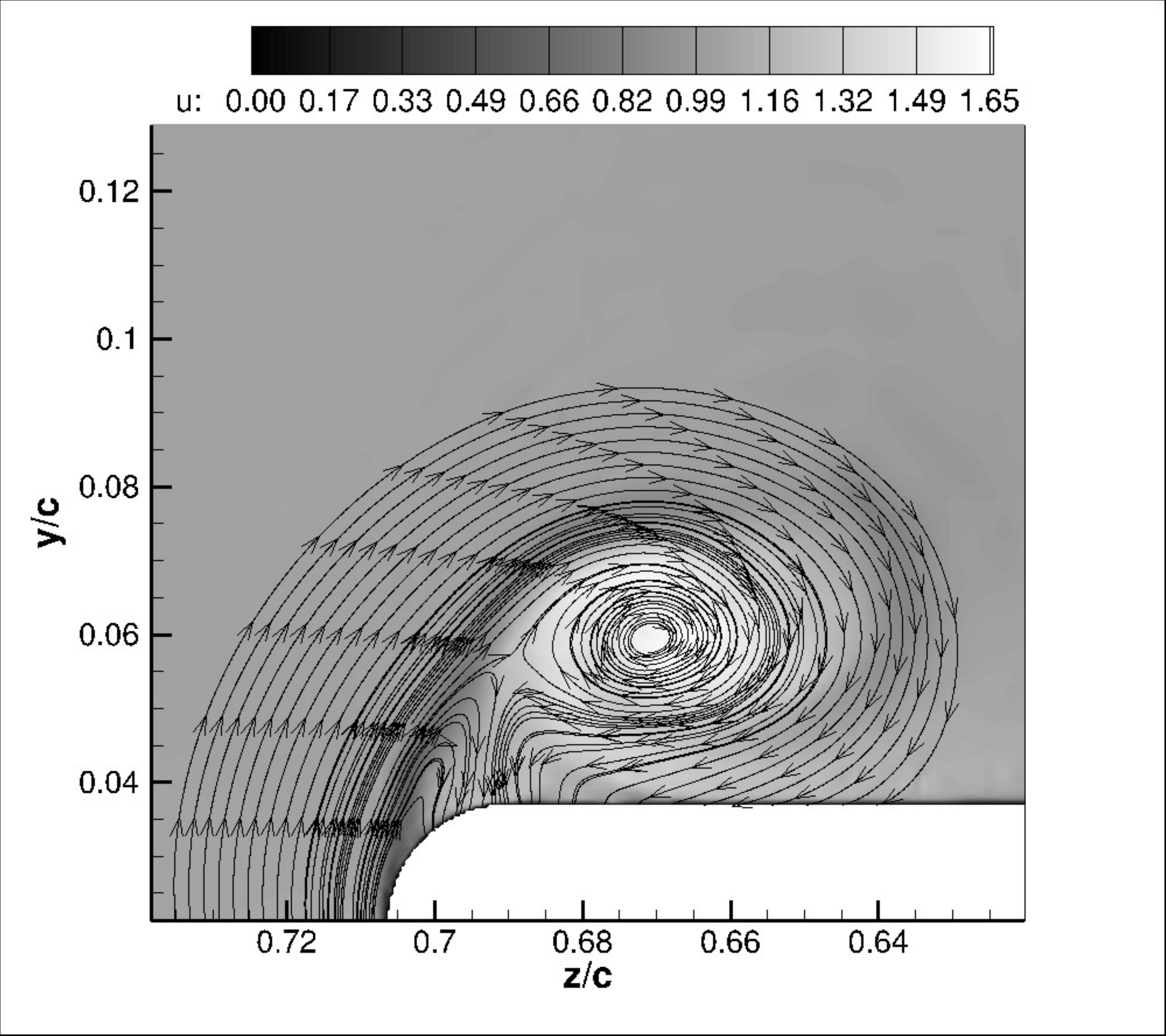}
              \put(20,70){(b)}
          \end{overpic}
\begin{overpic}[trim=20pt 20pt 20pt 20pt,clip,width=0.32\columnwidth]{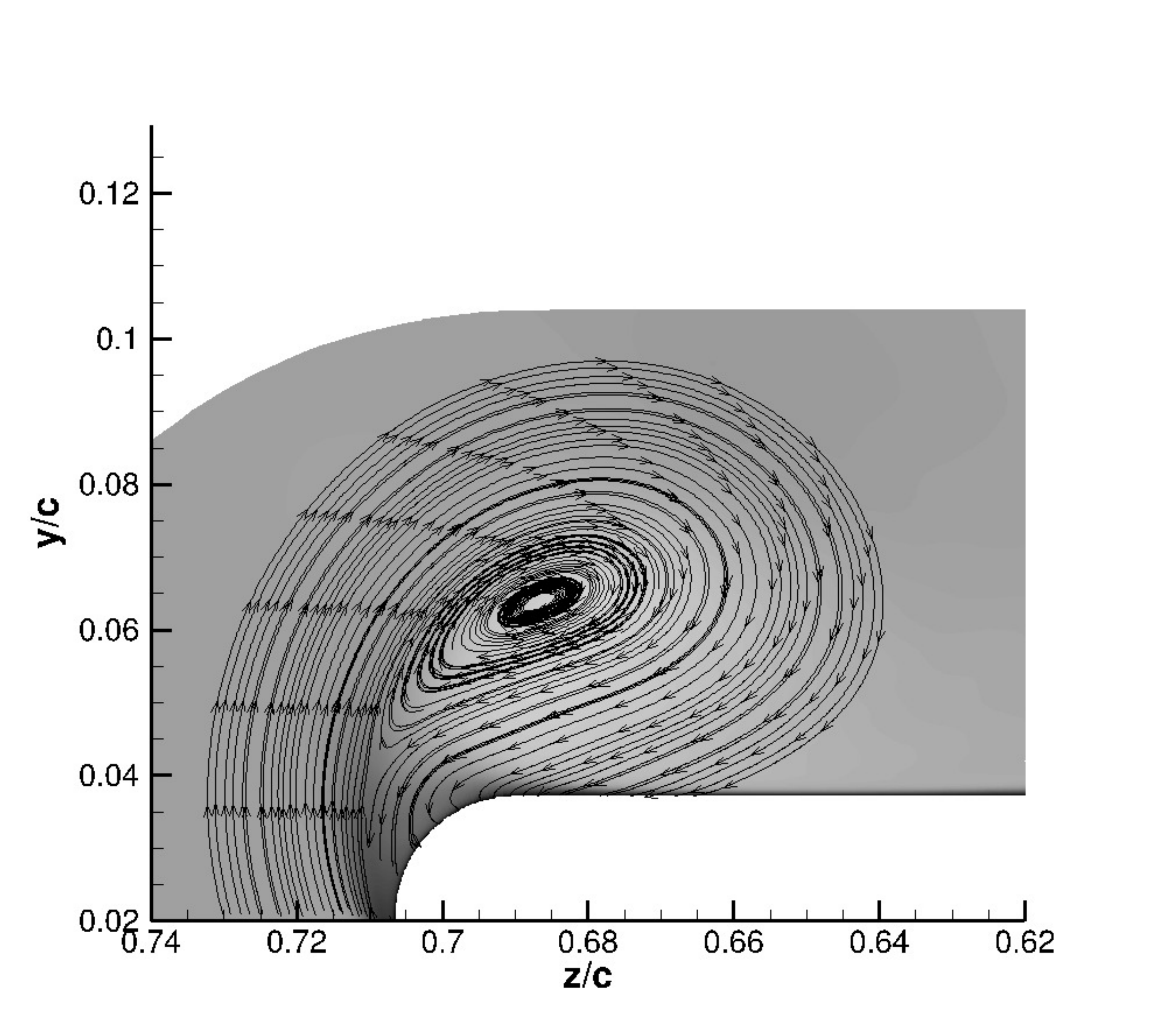}
              \put(20,70){(c)}
          \end{overpic}
    
    \caption{SVV-iLES computed streamlines and normalized time-averaged axial
      velocity at the crossflow plane $x/c=-0.115$ downstream of the trailing
      edge in b) compared against  experimental results from
      Chow \emph{et al.}\cite{chow}, in (a) and previous LES results by by Uzun \emph{et
        al.}\cite{Uzun:2006aa}, in (c).}
    \label{fig:9}
  \end{figure}

\new{The vertical location of the primary vortex core is computed to be 20\%
  lower than both the experiment and previous LES (Fig.~\ref{fig:8}a). The
  change in attitude as the vortex leaves the proximity of the wing surface
  follows a similar trend, with the core remaining at the same distance above
  the wing surface from streamwise location $x/c = -0.4$ from the trailing edge,
  to the trailing edge $x/c=0$ and then showing a pronounced upward trend from
  the leading edge to the experimentally reported $x/c=0.4$ location downstream
  of the leading edge. Fig.~\ref{fig:9}a shows the computed cross-section of the
  vertical vortex profile above the wing surface, with the same position
  measured experimentally (fig.~\ref{fig:9}b) and in the previous LES result of
  Uzun \emph{et al.} (fig~\ref{fig:9}c).  Despite a discrepancy regarding the
  vertical position of vortex core, we seem to be qualitatively agreeing with
  the experimental results. In particular, the shape of our computed vortex is
  closer to the isotropic round shape of the experimentally obtained vortex,
  particularly when compared to the previous LES result. The topology of the
  flow in the region $0.03<y/c<0.06$, with the presence of a secondary vortex,
  is also qualitatively closer to the experimental results than the previous
  LES.  We should note however that the location of the SVV-iLES computed
  structures is different with respect to those measured experimentally. Whilst
  it is difficult to explain this discrepancy without a further series of
  detailed simulations, one possible explanation is in the tangential grid
  spacing. We note that for computational reasons, this is clearly not as fine
  as the wall-normal direction, and this may therefore play a key role in
  influencing the detachment location and therefore vertical position of the
  vortex core. It is also interesting to remark how accurate the LES results
  from Uzun \emph{et al.} are at predicting the vertical position of the vortex
  core above the wing despite significantly different flow topologies
  (Fig.~\ref{fig:9}a and Fig.~\ref{fig:9}c).}


\subsubsection{Spanwise position of the vortex core.}

\begin{figure}[ht]
  \centering
\begin{overpic}[width=0.32\columnwidth]{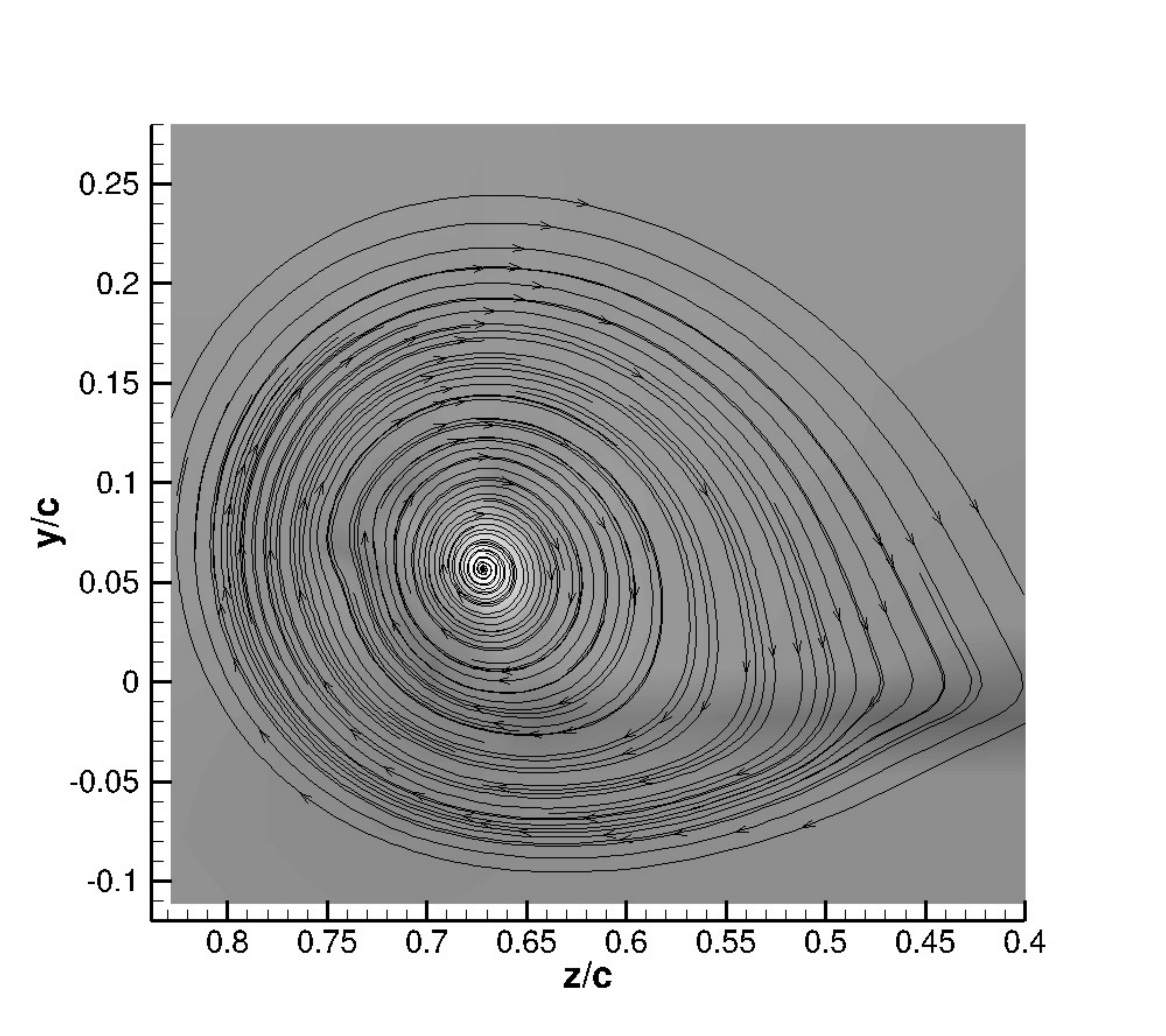}
              \put(20,70){(a)}
          \end{overpic}
\begin{overpic}[width=0.32\columnwidth]{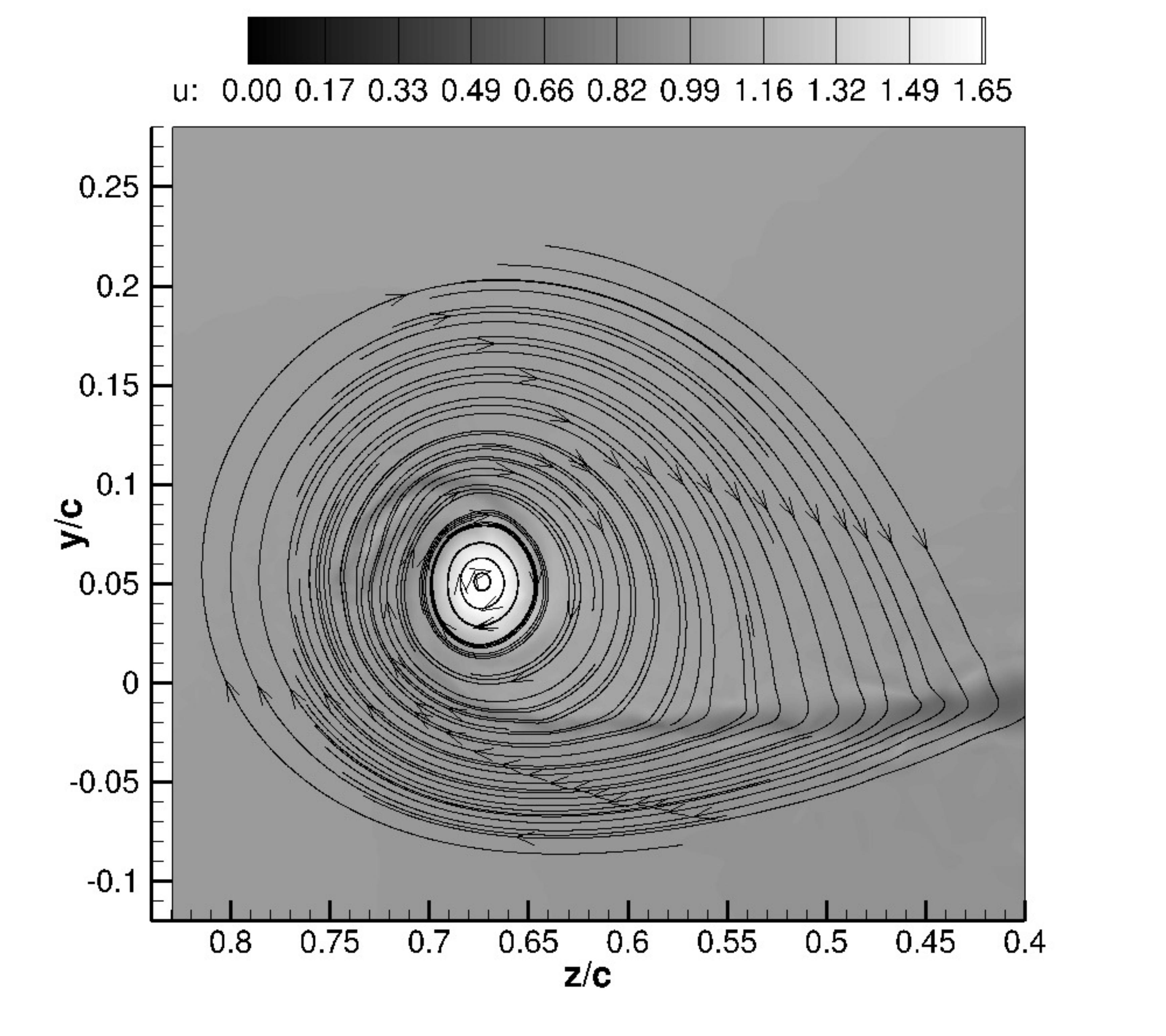}
              \put(20,70){(b)}
          \end{overpic}
\begin{overpic}[width=0.32\columnwidth]{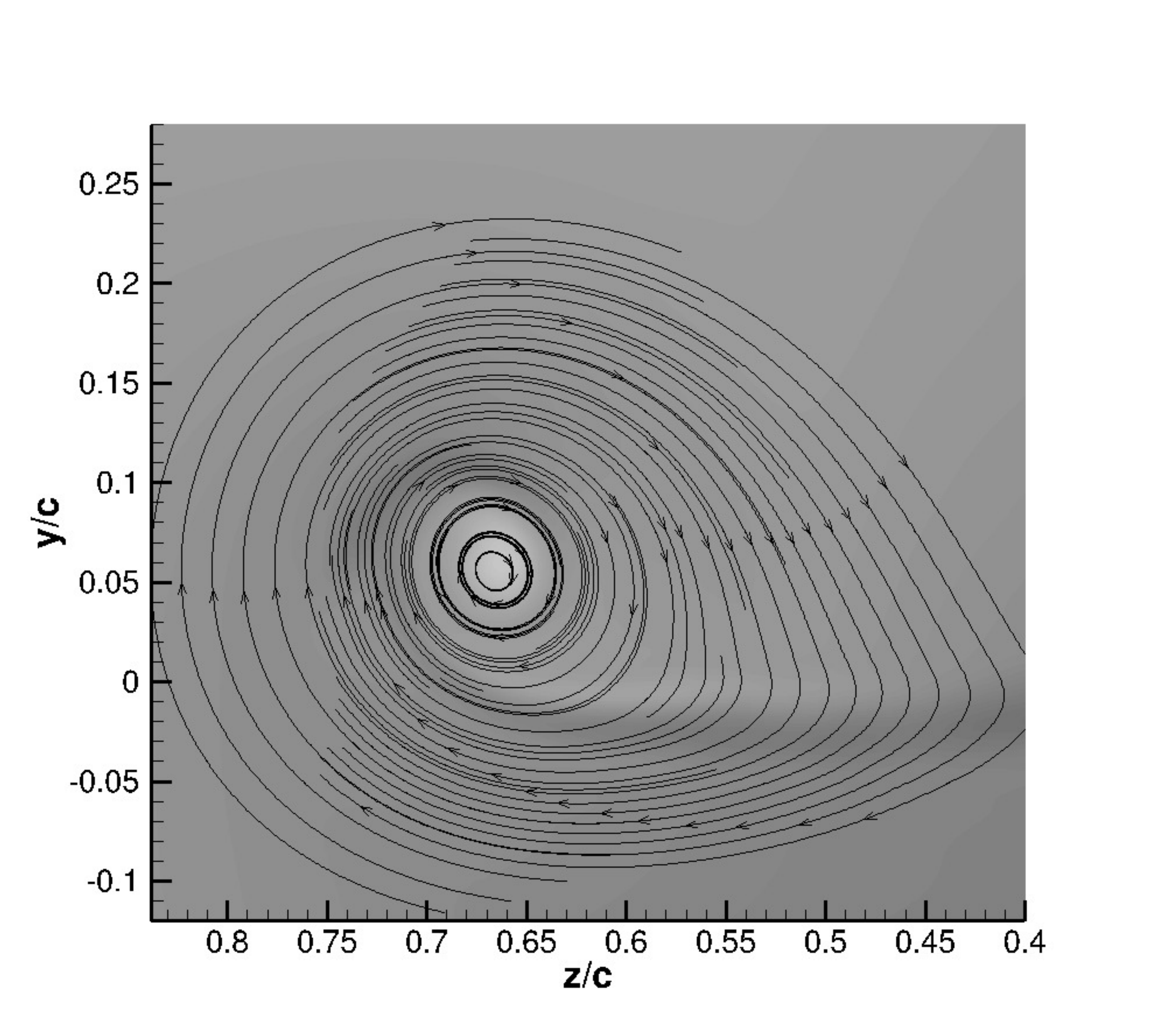}
              \put(20,70){(c)}
          \end{overpic}
  \caption{SVV-iLES computed streamlines and normalized time-averaged axial
    velocity at the crossflow plane $x/c=0.125$ downstream of the trailing edge
    in b) compared against experimental results from Chow \emph{et
      al.}~\cite{chow}, in (a) and previous LES results by by Uzun \emph{et
      al.}~\cite{Uzun:2006aa}, in (c).  Fig.~\ref{fig:7} acts as a companion
    figure to locate the $x/c=+0.125$ with respect to the primary and secondary
    vortices with a three dimensional perspective of the flow in this region.
    Fig.~\ref{fig:11} complements Fig.~\ref{fig:10}b in aiding the
    identification of the secondary vortex with respect to streamlines and also
    the streamwise component of the vorticity.}
        \label{fig:10}
\end{figure}

\begin{figure}[ht]
  \centering
\begin{overpic}[trim=20pt 20pt 20pt 20pt,clip,width=0.44\columnwidth]{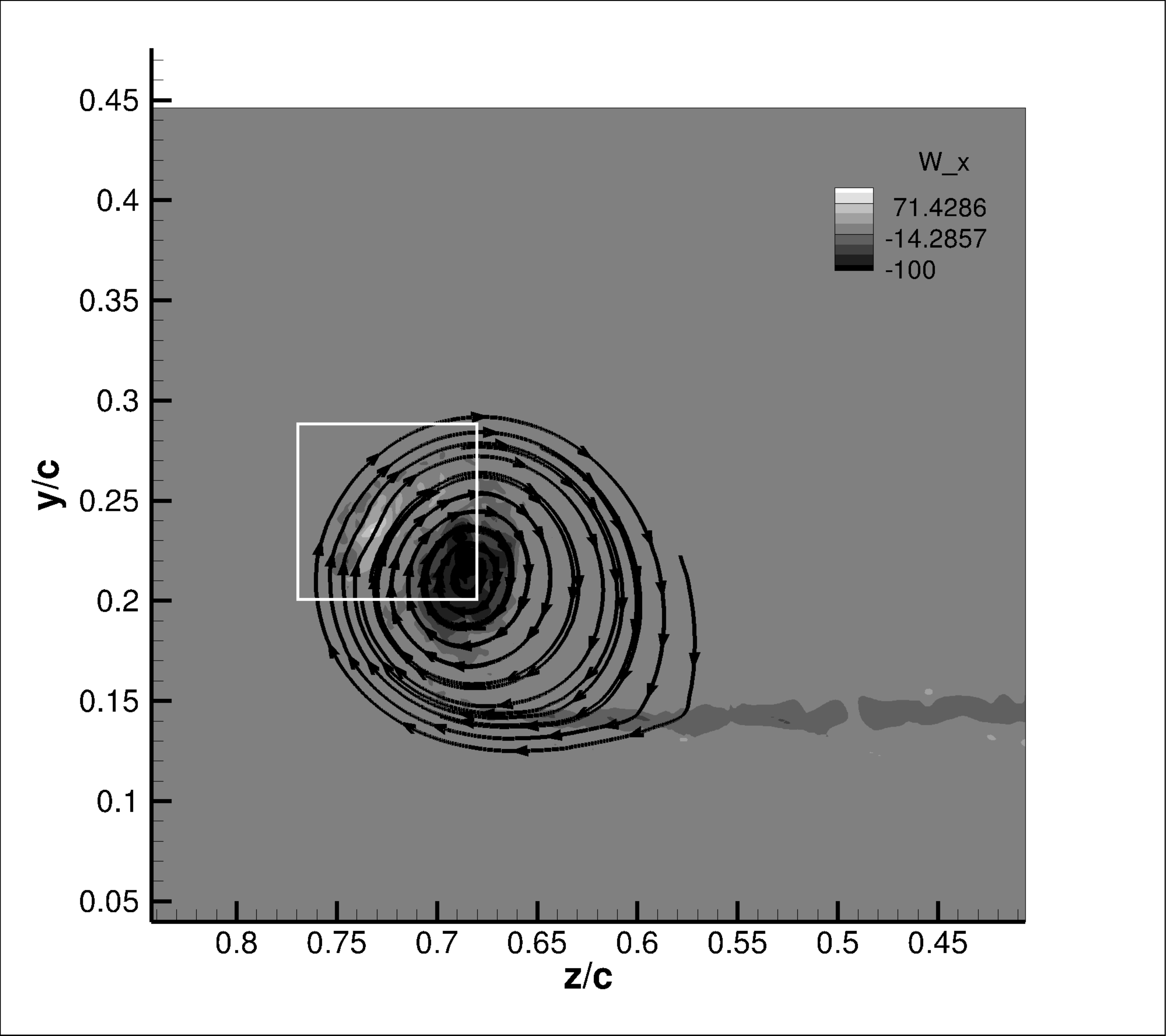}
              \put(20,70){(a)}
          \end{overpic}
\begin{overpic}[trim=20pt 20pt 20pt 20pt,clip,width=0.44\columnwidth]{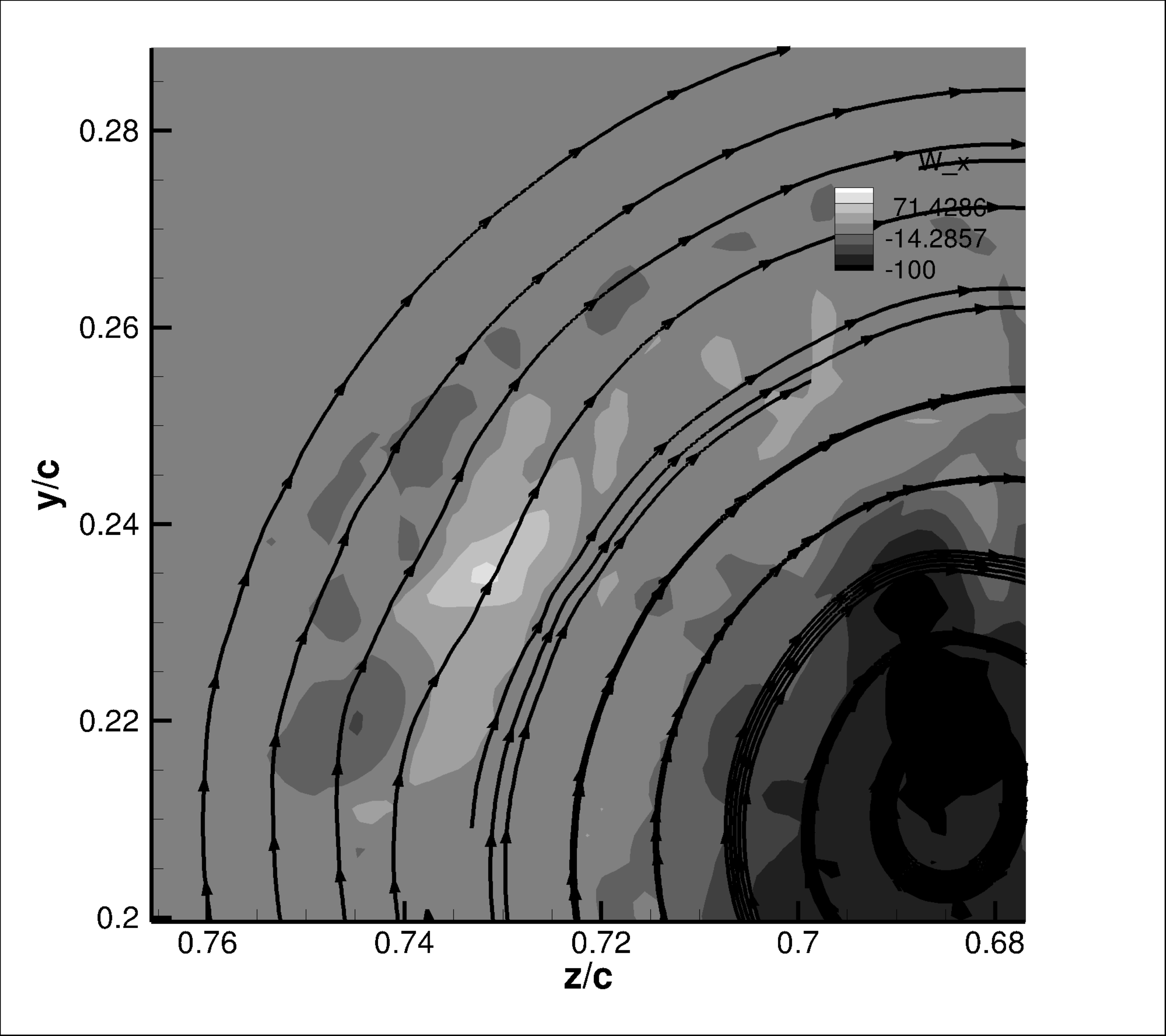}
              \put(20,70){(b)}
          \end{overpic}
  \caption{Streamwise component of the vorticity vector $\Omega_x$ at location
    $x/c=+0.125$. Both primary vortex, in dark grey, and secondary vortex in
    light grey in the II quadrant are visible with an overview of the flow in a)
    and the detail of the location of the secondary vortex in b). These figure
    acts as a companion to Fig.~\ref{fig:10}b.}
  \label{fig:11}
\end{figure}

\new{The spanwise position of the computed vortex core is compared against both
  the experimental data of Chow \emph{et al.}\cite{chow} and previous LES
  results by Uzun \emph{et al.}~\cite{Uzun:2006aa}.  Three key features of flow
  can be assessed by this figure: the location of the origin of the primary
  vortex, the location at the trailing edge and the evolution of the vortex as
  it leaves the vicinity of the surface of the wing. There is good agreement
  between experimental data and both LES regarding the position of the vortex at
  the trailing edge.  There is, however, a significant discrepancy regarding the
  origin of the vortex. Indeed both Chow \emph{et al.}~\cite{chow} (circles in
  Fig.~\ref{fig:8}b) and Uzun \emph{et al.}~\cite{Uzun:2006aa} (dot-dashed line
  in Fig.~\ref{fig:8}b) report the origin of the primary vortex in the region of
  the wing cap whereas the present results show the origin to be on the suction
  side of the wing around mid-chord. McAlister \&
  Takahashi~\cite{McAlister:1991} as well as Thompson~\cite{Thompson:1983}
  reported the origin of the vortex on the suction side of a NACA 0015 profile
  for a similar case.  McAlister \& Takahashi~\cite{McAlister:1991} also report
  tripping the boundary layer decreases the inboard movement of the primary
  vortex along the span without altering its distance above the wing. This may
  offer possible insight into the difference in spanwise trajectory between
  Uzun's LES computed vortex which has a more pronounced inboard movement than
  Chow's experiment.

  The most interesting feature of the flow that can be analysed with
  Fig.~\ref{fig:8}b is the evolution of the vortex as it leaves the vicinity of
  the surface of the wing. The experimental results of Chow \emph{et al.} show
  two distinct kinks at $x/c=0.125$ and then $x/c=0.452$. This is particularly
  evident when comparing against the previous LES of Uzun \emph{et al.}, where
  the progression of the vortex core moves steadily inboard. We believe these
  kinks are the result of the interaction between the primary vortex and a
  secondary vortex orbiting around it.  Evidence of this secondary vortex can be
  seen in both the comparisons of the streamwise-normal streamlines shown in
  Fig.~\ref{fig:10}a with the notable presence of a flat spot in the region
  $0.72<z/c<0.75$ and $0.01<y/c<0.07$. Our numerical results also show a similar
  flat spot, albeit in slightly different position, in region $0.72<z/c<0.75$
  and $0.06<y/c<0.1$. By plotting the streamwise component of the vorticity
  vector in this region we can correlate these flat spots to a weaker, counter
  rotating, secondary vortex (Fig.~\ref{fig:11}a). In Fig.~\ref{fig:11}b we
  enlarge this region of interest to evidence the weaker, counter rotating
  vortex in light grey. In these two figures the primary vortex appears in dark
  grey. This secondary, weaker, counter rotating vortex can also be seen when
  visualising the iso-helicity surfaces. In Fig.~\ref{fig:7} the primary vortex
  appears in grey and the secondary vortex in dark grey (or blue if visualised
  in color). The light grey (light blue) surface represents the spanwise
  location $x/c=+0.125$ at which the comparison is made between experiment,
  previous LES and present results for Fig.~\ref{fig:10}-\ref{fig:11}.  It is
  also interesting to note that the computed secondary vortex seems to be out of
  phase, in the streamwise direction, with respect to the experimentally
  computed vortex. Indeed it appears in the II quadrant whereas in the
  experimental results it appears in the III quadrant (when viewed as in
  Fig.~\ref{fig:10}-\ref{fig:11}).}


\section{Conclusion}
\label{sec:Conclusion}

The SVV-iLES workflow, developed for computing unsteady
vortex-dominated flows, has been assessed by comparing numerical results with
experimental data by Chow \emph{et al.}\cite{chow} for a NACA 0012 wingtip
vortex test case, at a higher Reynolds number than any LES study
\new{performed to date}.
\new{Overall, the results show the potential of this method to resolve the large
  scale features of the flow, without the use of explicit turbulence and
  sub-grid scale models. The use of an implicit LES presents a notable
  advantage over these methods, given that only two parameters are needed
  to control regularization and stability of the numerics.}

\new{Our results show better correlation with  experimental results than previous
numerical results,
  both in terms of the static pressure distribution, prediction of the jetting
  velocity, vortex spanwise location and the ability to resolve the
  secondary vortex interaction with the main wingtip vortex.}
In particular, we note that the results presented here show a good agreement in
the roll-up region where both the static pressure and velocity magnitude agree
within 10\% of experiment.
We also observe that although the mesh used in this study is coarser than ones
reported in other studies, as highlighted in Tab. \ref{t:scheme_comparison}, and
the Reynolds number is larger than previous investigations, the results of this
study demonstrate that the SVV-iLES method can still accurately capture the
essential features of the flow. \new{However we do note that unsteady simulation
  obviously requires significantly more compute resource as compared to steady
  RANS simulations.}  We \new{additionally} obtain a qualitatively good
agreement of the modeling of the vortex roll-up, and predict the secondary
vortex and its interaction with the primary vortex over the wingtip.
\new{We believe this can be attributed both to the lower diffusion and dispersion
  properties of the spectral/$hp$ element method and to the use of an
  isoparametric refinement technique which provides adequate wall-normal resolution
  in the sub-viscous layer of the boundary region.}

There are however some clear differences between the results presented here and
experimental data. It is clear that the suction peak on the wing appears further
downstream ($x/c=0.95$ instead of the experimental value of $x/c=0.85$). There
is also a visible difference in the location of the secondary vortex at location
$x/c=0.867$ downstream of the trailing edge.
\new{These points seem to indicate that despite successfully modeling the
  secondary vortex, the interaction between the primary and secondary vortex is
  not yet accurate enough to reproduce experimental results. The $p$-refinement
  study presented here shows that, whilst our results are well-resolved in terms
  of the polynomial space, a small increase in polynomial order does not
  generally lead to a better correlation with the experimental data. This is
  likely due an under-resolution of the wall-tangential directions across of the
  surface of the wing and in particular in the first $0.5c$, where the primary
  vortex originates. Therefore, whilst a large increase in polynomial order
  would hopefully yield a better correlation with the experimental results, a
  more efficient approach to achieve convergence is likely to be a combination
  of both mesh and $p$-refinement.  It is therefore clear that, together with
  improving the smoothness of the surface mesh and a further increase in
  Reynolds number to match the experiment, future studies should include
  additional local refinement in terms of element size in order to hopefully
  attain a closer agreement with the experimental results.

  In summary, the SVV-iLES method has been shown to be a compelling alternative
  for computing complex unsteady vortex dominated flows, such as the wingtip
  vortex, motivating its use for complex industrially relevant cases where
  high-fidelity computational fluid dynamics can become an enabling technology.}

\section*{Acknowledgments}
\new{The authors would like to thank Dr. Ali Uzun for sharing both experimental
  and LES results for figures~13-18. We also thank the reviewers of this
  manuscript for a number of helpful suggestions.} The authors acknowledge
support from the United Kingdom Turbulence Consortium (UKTC) under grant
EP/L000261/1 as well as from the Engineering and Physical Sciences Research
Council (EPSRC) for access to ARCHER UK National Supercomputing Service
(http://www.archer.ac.uk). DM acknowledges supported by the Laminar Flow Control
Centre funded by Airbus/EADS and EPSRC under grant EP/I037946. SJS additionally
acknowledges Royal Academy of Engineering support under their research chair
scheme. \new{We also acknowledge the support from the Imperial College London
  High Performance Computing facilities.}

\bibliographystyle{unsrt}
\bibliography{ref}

\begin{thebibliography}{10}

\bibitem{chow}
J.S. Chow, G.~Zilliac, and P.~Bradshaw.
\newblock Mean and turbulence measurements in the near field of a wingtip
  vortex.
\newblock {\em AIAA Journal}, 35(10):1561--1567, 1997.

\bibitem{Devenport:1995aa}
William~J. Devenport, Michael~C. Rife, Stergios~I. Liapis, and Gordon~J.
  Follin.
\newblock The structure and development of a wing-tip vortex.
\newblock {\em Journal of Fluid Mechanics}, 312:67--106, 4 1996.

\bibitem{Jacquin:2001aa}
L.~Jacquin, D.~Fabre, P.~Geffroy, and E.~Coustols.
\newblock {\em The properties of a transport aircraft wake in the extended near
  field - An experimental study}.
\newblock American Institute of Aeronautics and Astronautics, 2001.

\bibitem{Heyes:2004}
A~Heyes, R~Jones, and D~Smith.
\newblock Wandering of wingtip vortices.
\newblock In {\em 12th Internation symposium on application of laser techniques
  to fluid mechanics}, 2004.

\bibitem{Uzun:2006aa}
Ali Uzun, M.~Yousuff Hussaini, and Craig~L. Streett.
\newblock Large-eddy simulation of a wing tip vortex on overset grids.
\newblock {\em AIAA Journal}, 44(6):1229--1242, 2006.

\bibitem{Uzun:2010aa}
Ali Uzun and M.~Yousuff Hussaini.
\newblock Simulations of vortex formation around a blunt wing tip.
\newblock {\em AIAA Journal}, 48(6):1221--1234, 2010.

\bibitem{Rossow1999507}
Vernon~J. Rossow.
\newblock Lift-generated vortex wakes of subsonic transport aircraft.
\newblock {\em Progress in Aerospace Sciences}, 35(6):507 -- 660, 1999.

\bibitem{Ghias:2005aa}
Reza Ghias, Rajat Mittal, Haibo Dong, and Thomas Lund.
\newblock {\em Study of Tip-Vortex Formation Using Large-Eddy Simulation}.
\newblock American Institute of Aeronautics and Astronautics, 2005.

\bibitem{Dacles-Mariani:1995aa}
Jennifer Dacles-Mariani, Gregory~G. Zilliac, Jim~S. Chow, and Peter Bradshaw.
\newblock Numerical/experimental study of a wingtip vortex in the near field.
\newblock {\em AIAA Journal}, 33(9):1561--1568, 1995.

\bibitem{Spalart:1998}
Philippe~R. Spalart.
\newblock Airplane trailing vortices.
\newblock {\em Annual Review of Fluid Mechanics}, 30(1):107--138, 1998.

\bibitem{Churchfield:2013aa}
Matthew~J. Churchfield and Gregory~A. Blaisdell.
\newblock Reynolds stress relaxation turbulence modeling applied to a wingtip
  vortex flow.
\newblock {\em AIAA Journal}, 51(11):2643--2655, 2013.

\bibitem{Satti:2012aa}
Rajani Satti, Yanbing Li, Richard Shock, and Swen Noelting.
\newblock Unsteady flow analysis of a multi-element airfoil using lattice
  boltzmann method.
\newblock {\em AIAA Journal}, 50(9):1805--1816, 2012.

\bibitem{Sagaut:2001}
Pierre Sagaut.
\newblock {\em Large Eddy Simulation for Incompressible Flows}.
\newblock Springer Berlin Heidelberg, 2001.

\bibitem{bolis}
A.~Bolis, C.~D. Cantwell, R.~M. Kirby, and S.~J. Sherwin.
\newblock From h to p efficiently: optimal implementation strategies for
  explicit time-dependent problems using the spectral/hp element method.
\newblock {\em International Journal for Numerical Methods in Fluids},
  75(8):591--607, 2014.

\bibitem{Churchfield:2011aa}
Matthew Churchfield and Gregory Blaisdell.
\newblock {\em A Reynolds Stress Relaxation Turbulence Model Applied to A
  Wingtip Vortex Flow}.
\newblock American Institute of Aeronautics and Astronautics, 2011.

\bibitem{Dacles-Mariani:1996aa}
Jennifer Dacles-Mariani, Dochan Kwak, and Gregory Zilliac.
\newblock {\em Accuracy assessment of a wingtip vortex flowfield in the
  near-field region}.
\newblock American Institute of Aeronautics and Astronautics, 1996.

\bibitem{Duraisamy:2006aa}
Karthikeyan Duraisamy and James~D. Baeder.
\newblock Numerical simulation of the effects of spanwise blowing on wing-tip
  vortex formation and evolution.
\newblock {\em Journal of Aircraft}, 43(4):996--1006, 2006.

\bibitem{Satti:2011aa}
Rajani Satti, Yanbing Li, Richard Shock, and Brad Duncan.
\newblock {\em Computational Analysis of a Wingtip Vortex in the Near-Field
  using LBM-VLES Approach}.
\newblock American Institute of Aeronautics and Astronautics, 2011.

\bibitem{Tadmor:1989aa}
Eitan Tadmor.
\newblock Convergence of spectral methods for nonlinear conservation laws.
\newblock {\em SIAM Journal on Numerical Analysis}, 26(1):30--44, 02 1989.

\bibitem{Green:1991}
S.~I. Green and A.~J. Acosta.
\newblock Unsteady flow in trailing vortices.
\newblock {\em Journal of Fluid Mechanics}, 227:107--134, 6 1991.

\bibitem{giuni:2013}
M.~Giuni.
\newblock {\em Formation and early development of wingtip vortices}.
\newblock PhD thesis, University of Glasgow, 2013.

\bibitem{Giuni:2013aa}
Michea Giuni and Richard~B. Green.
\newblock Vortex formation on squared and rounded tip.
\newblock {\em Aerospace Science and Technology}, 29(1):191--199, 8 2013.

\bibitem{Giuni:2011aa}
Michea Giuni and Emmanuel Benard.
\newblock {\em Analytical/Experimental Comparison of the Axial Velocity in
  Trailing Vortices}.
\newblock American Institute of Aeronautics and Astronautics, 2011.

\bibitem{Fabre:2000}
David Fabre and Laurent Jacquin.
\newblock Stability of a four-vortex aircraft wake model.
\newblock {\em Physics of Fluids (1994-present)}, 12(10):2438--2443, 2000.

\bibitem{Fabre:2002}
D.~Fabre, L.~Jacquin, and A.~Loof.
\newblock Optimal perturbations in a four-vortex aircraft wake in
  counter-rotating configuration.
\newblock {\em Journal of Fluid Mechanics}, 451:319--328, 1 2002.

\bibitem{Dieterle:1999}
L.~Dieterle, K.~Ehrenfried, R.~Stuff, G.~Schneider, P.~Coton, J.C. Monnier, and
  J.F. Lozier.
\newblock Quantitative flow field measurements in a catapult facility using
  particle image velocimetry.
\newblock In {\em Instrumentation in Aerospace Simulation Facilities, 1999.
  ICIASF 99. 18th International Congress on}, pages 1/1--110, 1999.

\bibitem{zuhal:2001}
L.R. Zuhal.
\newblock {\em Formation and Near-field Dynamics of a Wingtip Vortex}.
\newblock PhD thesis, California Institute of Technology, 2001.

\bibitem{Zuhal:2001aa}
Lavi Zuhal and Morteza Gharib.
\newblock {\em Near field dynamics of wing tip vortices}.
\newblock American Institute of Aeronautics and Astronautics, 2001.

\bibitem{McAlister:1991}
K.~W. McAlister and R.~K. Takahashi.
\newblock Avscom technical report 91-a-003 naca 0015 wing pressure and trailing
  vortex measurements.
\newblock Technical report, Nasa Tecnical Paper 3151, 1991.

\bibitem{Churchfield:2009ab}
Matthew~J. Churchfield and Gregory~A. Blaisdell.
\newblock Numerical simulations of a wingtip vortex in the near field.
\newblock {\em Journal of Aircraft}, 46(1):230--243, 2009.

\bibitem{DACLES-MARIANI:1993aa}
J.~Dacles-Mariani, S.~Rogers, D.~Kwak, G.~Zilliac, and J.~Chow.
\newblock {\em A computational study of wingtip vortex flowfield}.
\newblock American Institute of Aeronautics and Astronautics, 1993.

\bibitem{Craft2006684}
T.J. Craft, A.V. Gerasimov, B.E. Launder, and C.M.E. Robinson.
\newblock A computational study of the near-field generation and decay of
  wingtip vortices.
\newblock {\em International Journal of Heat and Fluid Flow}, 27(4):684 -- 695,
  2006.
\newblock Special Issue of The Fourth International Symposium on Turbulence and
  Shear Flow Phenomena - 2005 Special Issue of The Fourth International
  Symposium on Turbulence and Shear Flow Phenomena - 2005.

\bibitem{Duraisamy:2005}
G.~Iaccarino K.~Duraisamy.
\newblock Curvature correction and application of the v2 − f turbulence model
  to tip vortex flows.
\newblock Technical report, Center for Turbulence Research - Annual Research
  Briefs, Stanford, CA, 2005.

\bibitem{Churchfield:2008aa}
Matthew Churchfield and Gregory Blaisdell.
\newblock {\em The Lag RST Turbulence Model Applied to a Vortex Flow}.
\newblock American Institute of Aeronautics and Astronautics, 2008.

\bibitem{Churchfield:2009aa}
Matthew~J. Churchfield and Gregory~A. Blaisdell.
\newblock Numerical simulations of a wingtip vortex in the near field.
\newblock {\em Journal of Aircraft}, 46(1):230--243, 2009.

\bibitem{Fleig:2004aa}
Oliver Fleig, Makoto Iida, and Chuichi Arakawa.
\newblock Wind turbine blade tip flow and noise prediction by large-eddy
  simulation.
\newblock {\em Journal of Solar Energy Engineering}, 126(4):1017--1024, 11
  2004.

\bibitem{Jiang:2008}
Li~Jiang, Jiangang Cai, and Chaoqun Liu.
\newblock Large-eddy simulation of wing tip vortex in the near field.
\newblock {\em International Journal of Computational Fluid Dynamics},
  22(5):289--330, 2008.

\bibitem{Menter:1994aa}
F.~R. Menter.
\newblock Two-equation eddy-viscosity turbulence models for engineering
  applications.
\newblock {\em AIAA Journal}, 32(8):1598--1605, 2015/07/18 1994.

\bibitem{Cantwell:2014}
C.~D. Cantwell, D.~Moxey, A.~Comerford, A.~Bolis, G.~Rocco, G.~Mengaldo,
  D.~de~Grazia, S.~Yakovlev, J.-E.~W. Lombard, D.~Ekelschot, B.~Jordi,
  Y.~Mohamied, C.~Eskilsson, B.~Nelson, P.~Vos, C.~Biotto, R.~M. Kirby, and
  S.~J. Sherwin.
\newblock {Nektar++: An open-source spectral/hp element framework. Accepted for
  publication in Computer Physics Communications}.
\newblock 2014.

\bibitem{spencer}
G.~Karniadakis and S.~Sherwin.
\newblock {\em Spectral/hp Element Methods for Computational Fluid Dynamics}.
\newblock Oxford University Press, second edition, 2005.

\bibitem{Splitting}
G.~Karniadakis, M.~Israeli, and S.~Orszag.
\newblock High-order splitting methods for incopressible navier-stokes
  equations.
\newblock {\em J. Comp. Phys}, 97:414--443, 1991.

\bibitem{Mengaldo:2014}
G.~Mengaldo, D.~De Grazia, D.~Moxey, P.E. Vincent, and S.J. Sherwin.
\newblock Dealiasing techniques for high-order spectral element methods on
  regular and irregular grids.
\newblock {\em Journal of Computational Physics}, 2014.

\bibitem{Kirby20063128}
Robert~M. Kirby and Spencer~J. Sherwin.
\newblock Stabilisation of spectral/hp element methods through spectral
  vanishing viscosity: Application to fluid mechanics modelling.
\newblock {\em Computer Methods in Applied Mechanics and Engineering},
  195(23--24):3128 -- 3144, 2006.
\newblock Incompressible \{CFD\}.

\bibitem{Karamanos:2000}
G.-S. Karamanos and G.~E. Karniadakis.
\newblock A spectral vanishing viscosity method for large-eddy simulations.
\newblock {\em J. Comput. Phys.}, 163(1):22--50, September 2000.

\bibitem{Pasquetti:2005}
R.~Pasquetti.
\newblock Spectral vanishing viscosity method for les: sensitivity to the svv
  control parameters.
\newblock {\em Journal of Turbulence}, 6, 2005.

\bibitem{Severac:2007}
E.~Severac and E.~Serre.
\newblock A spectral vanishing viscosity for the les of turbulent flows within
  rotating cavities.
\newblock {\em J. Comput. Phys.}, 226(2):1234--1255, October 2007.

\bibitem{Xu:2006aa}
Chuanju Xu.
\newblock Stabilization methods for spectral element computations of
  incompressible flows.
\newblock {\em Journal of Scientific Computing}, 27(1-3):495--505, 2006.

\bibitem{Lamballais20113270}
Eric Lamballais, V{\'e}ronique Fortun{\'e}, and Sylvain Laizet.
\newblock Straightforward high-order numerical dissipation via the viscous term
  for direct and large eddy simulation.
\newblock {\em Journal of Computational Physics}, 230(9):3270 -- 3275, 2011.

\bibitem{lowenergy}
S.J. Sherwin and M.~Casarin.
\newblock Low-energy basis preconditioning for elliptic substructured solvers
  based on unstructured spectral/hp element discretization.
\newblock {\em J. Comp. Phys}, 171:394--417, 2001.

\bibitem{Kirby:2002aa}
Robert~M. Kirby and George~Em Karniadakis.
\newblock Coarse resolution turbulence simulations with spectral vanishing
  viscosity---large-eddy simulations (svv-les).
\newblock {\em Journal of Fluids Engineering}, 124(4):886--891, 12 2002.

\bibitem{Moura2015}
R.C. Moura, S.J. Sherwin, and J.~Peir{\'o}.
\newblock Linear dispersion-diffusion analysis and its application to
  under-resolved turbulence simulations using discontinuous galerkin
  spectral/hp methods.
\newblock {\em Journal of Computational Physics}, 2015.

\bibitem{Pasquetti:2006aa}
Richard Pasquetti.
\newblock Spectral vanishing viscosity method for large-eddy simulation of
  turbulent flows.
\newblock {\em Journal of Scientific Computing}, 27(1-3):365--375, 2006.

\bibitem{Pasquetti:2008aa}
R.~Pasquetti, E.~S{\'e}verac, E.~Serre, P.~Bontoux, and M.~Sch{\"a}fer.
\newblock From stratified wakes to rotor--stator flows by an svv--les method.
\newblock {\em Theoretical and Computational Fluid Dynamics}, 22(3-4):261--273,
  2008.

\bibitem{volino-1998}
Pascal Volino and N~Magenat Thalmann.
\newblock The spherigon: a simple polygon patch for smoothing quickly your
  polygonal meshes.
\newblock In {\em Computer Animation 98. Proceedings}, pages 72--78. IEEE,
  1998.

\bibitem{Moxey:2014}
D.~Moxey, M.~Hazan, J.~Peir\'o, and S.~J. Sherwin.
\newblock {An isoparametric approach to high-order curvilinear boundary-layer
  meshing}.
\newblock to appear in Comput. Meth. Appl. Mech. Eng., sep 2014.

\bibitem{Dacles-Mariani:1997aa}
Jennifer Dacles-Mariani, Mohamed Hafez, and Dochan Kwak.
\newblock {\em Prediction of wake-vortex flow in the near- and
  intermediate-fields behind wings}.
\newblock American Institute of Aeronautics and Astronautics, 1997.

\bibitem{Dong201483}
S.~Dong, G.E. Karniadakis, and C.~Chryssostomidis.
\newblock A robust and accurate outflow boundary condition for incompressible
  flow simulations on severely-truncated unbounded domains.
\newblock {\em Journal of Computational Physics}, 261(0):83 -- 105, 2014.

\bibitem{Karniadakis1991}
G.~E. Karniadakis, M.~Israeli, and Orszag~S. A.
\newblock High-order splitting methods for the incompressible navier-stokes
  equations.
\newblock {\em Journal of Computational Physics}, 97(2):414 -- 443, 1991.

\bibitem{Bradshaw:1973}
P.~Bradshaw.
\newblock Effects of streamline curvature on turbulent flow.
\newblock Technical report, AGARD-AG-169, 1973.

\bibitem{Thompson:1983}
D.H. Thompson.
\newblock Aerodynamics note 421 : A flow visualization study of tip vortex
  formation.
\newblock Technical report, Defence Science and Technology Organisation
  Aeronautical Research Laboratories, Melbourne Australia, 1983.

\end{thebibliography}

\end{document}